\documentclass[onecolumn,journal]{IEEEtran}
\usepackage{times}
\usepackage{cite}
\usepackage{color}
\usepackage{epsf}
\usepackage{epsfig}
\usepackage{graphics}
\usepackage{epstopdf}
\usepackage[small]{caption}
\usepackage{amsmath}
\usepackage{amssymb}
\usepackage{amsthm}
\usepackage{amsxtra}
\usepackage[ruled]{algorithm2e}
\usepackage{algorithmic}
\usepackage{enumerate}
\usepackage{multirow}
\usepackage{enumerate}
\usepackage{amssymb}
\usepackage{lipsum}
\usepackage{setspace}
\usepackage{adjustbox}
\usepackage{multirow}
\usepackage{subcaption}
\usepackage{url}
\usepackage{color,soul}
\usepackage{geometry}
\usepackage[utf8]{inputenc}
\usepackage{url}
\usepackage{array}
\usepackage{makecell}
\usepackage[normalem]{ulem}

\usepackage{rotating} 
\usepackage{graphicx} 

\newcommand{\Rmnum}[1]{\expandafter\@slowromancap\romannumeral #1@}

\makeatother
\begin{document}
\title{Peer-to-Peer Energy Systems for Connected Communities: A Review of Recent Advances and Emerging Challenges}
\author{Wayes Tushar$^\text{a}$, Chau Yuen$^{\text{b,*}}$, Tapan K. Saha$^\text{a}$, Thomas Morstyn$^\text{c}$, Archie Chapman$^\text{a}$, M. Jan E. Alam$^\text{d}$, Sarmad Hanif$^\text{d}$, and H. Vincent Poor$^\text{e}$\\$^\text{a}$The University of Queensland, Brisbane, Australia\\$^\text{b}$Singapore University of Technology and Design, Singapore\\$^\text{c}$University of Edinburgh, United Kingdom\\$^\text{d}$Pacific Northwest National Laboratory, WA, USA\\$^\text{e}$Princeton University, NJ, USA
\thanks{$^*$Corresponding author at: Engineering Product Development Pillar, Singapore University of Technology and Design, 8 Somapah Road, Singapore 487372.}
\thanks{\emph{E-mail addresses:} w.tushar@uq.edu.au (W. Tushar), yuenchau@sutd.edu.sg (C. Yuen), saha@itee.uq.edu.au (T. K. Saha), thomas.morstyn@ed.ac.uk (T. Morstyn), archie.chapman@uq.edu.au (A. Chapman), mdjane.alam@pnnl.gov (M. J. E. Alam), sarmad.hanif@pnnl.gov (S. Hanif), poor@princeton.edu (H. V. Poor).}}
\IEEEoverridecommandlockouts
\maketitle
\doublespace
\begin{abstract}
After a century of relative stability of the electricity industry, extensive deployment of distributed energy resources and recent advances in computation and communication technologies have changed the nature of how we consume, trade, and apply energy. The power system is facing a transition from its traditional hierarchical structure to a more deregulated model by introducing new energy distribution models such as peer-to-peer (P2P) sharing for connected communities. The proven effectiveness of P2P sharing in benefiting both prosumers and the grid has been demonstrated in many studies and pilot projects. However, there is still no extensive implementation of such sharing models in today's electricity markets. This paper aims to shed some light on this gap through a comprehensive overview of recent advances in the P2P energy system and an insightful discussion of the challenges that need to be addressed in order to establish P2P sharing as a viable energy management option in today's electricity market. To this end, in this article, we provide some background on different aspects of P2P sharing. Then, we discuss advances in P2P sharing through a systematic domain-based classification. We also review different pilot projects on P2P sharing across the globe. Finally, we identify and discuss a number of challenges that need to be addressed for scaling up P2P sharing in the electricity market followed by concluding remarks at the end of the paper.
\end{abstract}
\begin{IEEEkeywords}
Peer-to-peer network, energy sharing, connected community, negawatt, review paper, electric vehicle, solar, storage, renewable.
\end{IEEEkeywords}
\section{Introduction}\label{sec:introduction}Energy systems are undergoing a rapid transition to accommodate the increasing penetration of embedded distributed energy resources (DER), such as solar photovoltaic (PV) arrays and wind turbines. In Australia, for example, 2 GW of installed rooftop PV capacity has been installed as of December 2019~\cite{SolarReport2020}, which is expected to increase up to 25 GW by 2030~\cite{SolarPV2030}. This extensive integration of DER to the energy network opens opportunities to provide values for both the grid and the DER owners. From a grid perspective, on the one hand, DER can benefit by providing flexibility to improve localized network performance issues such as voltage fluctuation~\cite{Xu_TPWRS_May_2020} and network management capacity \cite{Scott_TSG_Nov_2019}. On the other hand, prosumers can reduce their energy cost by using on-site generation from their DER and make revenues by sharing the surplus energy~\cite{Luth_AE_Nov_2018}. 

Such capacity of energy sharing in the local energy market makes the DER owners active prosumers~\cite{Thomas_Nature_2018} - energy consumers who also produce energy from their DER. However, the benefit that a prosumer can reap by trading its energy in the market could be marginal if it cannot decide on its energy trading parameters, such as the how much energy to share and the price per unit of energy, independently~\cite{Tushar_TIE_Apr_2015}. Given this context, peer-to-peer (P2P) energy sharing has emerged as a platform that can facilitate the independent decision-making process of prosumers to trade their energy within a connected community~\cite{Tushar_AE_June_2019}. In P2P sharing, a prosumer can independently decide on its energy sharing parameters such as how much energy to share and the price and determine who to share the energy with and when to share. Here, it is important to note that, in P2P trading, although a centralized controller or a third party may partially influence the decision-making process of a prosumer, it cannot directly control what a prosumer chooses to trade with other community members. For example, a third party or centralized controller may impose a constraint on the maximum power injection limit for a prosumer in the P2P market \cite{Imran_PESGM_2020}, which will influence the decision of the prosumer, but how much energy the prosumer will trade with other prosumers within the community given that injection limit is decided by the prosumer independently without any direct control from the third party (or centralized controller).  The objective of P2P trading also include 1) reducing greenhouse gas emission, 2) enabling consumers without DER to participate in low-cost energy trading, 3) providing demand flexibility and energy services to the grid, and 4) ensuring greater prosumer privacy. Moreover, the advancement in controllable DER techniques \cite{Arnold_TPWRS_Jan_2018} to control the power injection limit of prosumers has further motivated the evolution of P2P trading with the promise to not violate network constraints during energy trading.

Consequently, there has been a growing interest in P2P energy sharing research for the last few years. The focuses of these research can be generally divided into three categories: 1) research that focuses on the decision-making process of different participating prosumers with the aim to achieve some targeted performance improvements either in the individual or in the community level~\cite{Tushar_TSG_Jan_2020}; 2) research that addresses the impact of P2P sharing on the physical energy network~\cite{Archie_P2P_Sept_2019,Zhang_TSG_July_2020}, and 3) research that studies the development of platforms to enable P2P sharing~\cite{Siano_Systems_Sept_2019}. Further, a fair number of pilot projects on P2P sharing and relevant energy management techniques are also being established in different parts of the world~\cite{Zhang_EP_May_2017}. Interestingly, despite these extensive efforts, to date, there has been no consideration of pathways to implementing P2P sharing models in today's electricity market. One potential reason could be the lack of a comprehensive understanding of the gap between what has been developed to date and what else needs to be done for scaling up P2P sharing.

Given this context, in this paper, we aim to address this gap by shedding some light on potential barriers to implementing P2P sharing in existing electricity market frameworks and regulatory regimes. We do so by providing a comprehensive overview of recent advances in the P2P energy system and an insightful discussion of the challenges that need to be further addressed to establish P2P sharing as a viable energy management option in today's electricity market through following contributions: 
\begin{itemize}
\item We provide a detailed background of different aspects of the P2P energy sharing system. 
\item We discuss advances in P2P sharing through a systematic domain-based classification. 
\item We review different pilot projects on P2P sharing and relevant energy management technologies across the globe. 
\item We identify and discuss a number of challenges that need to be addressed for scaling up P2P sharing in the electricity market. 
\end{itemize}

We note that several recent survey papers have also contributed extensively to the body of energy sharing knowledge. Most of these studies focused on very specific topics such as blockchain~\cite{Andoni_RSER_Feb_2019}, distributed ledger~\cite{Siano_Systems_Sept_2019}, game theory~\cite{Tushar_SPM_July_2018}, computational approaches~\cite{Juhar_Energies_June_2018}, and markets~\cite{Sousa_RSER_Apr_2019}. However, due to the lack of a general overview of the topic, their capacities in identifying further modifications that are needed to prepare P2P sharing as a viable energy management options in the current electricity market are rather limited. Two further reviews require additional comments. First, the authors in \cite{Tushar_TSG_Jan_2020} provided a comprehensive general overview of various challenges addressed by existing studies in P2P trading. Nevertheless, the discussion in \cite{Tushar_TSG_Jan_2020} revolved around different challenges, rather than different domains of the P2P energy system. This makes it difficult to identify the developments and subsequent gaps in various domains of the P2P energy system from the discussion in \cite{Tushar_TSG_Jan_2020}. Further, an overview of existing pilot projects is missing in the study. In this work, we address this issue by choosing a domain-based classification, while surveying existing studies as well as a providing comprehensive overview of existing pilots on P2P sharing and relevant energy trading markets across the globe. The proposed study is also different from existing reviews in terms of contents, organization, and the focus on discussion. 

Second, \cite{Guerrero_RSER_Oct_2020} consider coordination and optimization methods for facilitating the integration of small-scale DER into low- and medium-voltage networks. The authors in \cite{Guerrero_RSER_Oct_2020} focus on three general approaches to coordinating prosumers: (i) uncoordinated approaches that only consider energy management of an individual user; (ii) approaches that cast the coordinated energy management problem as an optimisation problem; and (iii) peer-to-peer energy trading. Although \cite{Guerrero_RSER_Oct_2020} investigate which integration methods can be implemented with different levels of network awareness and their capability to address network or consumer interests, their focus is on computational mechanisms, so they ignore many of the additional implementation challenges brought to light in the current work.  Similarly, the case studies in \cite{Guerrero_RSER_Oct_2020} focus on energy trading between simple prosumers, and although they rigorously examine the different computational approaches to coordinating prosumers, they do not discuss advances in controllable DER technology that facilitating the participation of new domains in P2P energy trading, which are considered in this survey.

The rest of the paper is organized as follows. We provide background on several topics related to P2P energy sharing in Section \ref{sec:background}. In Section \ref{sec:AdvancementofP2P}, a detailed analysis of recent advancement in P2P energy sharing is given based on a systematic domain-based classification, followed by a detailed discussion on various P2P sharing projects around the world in Section \ref{sec:PilotProjects}. We summarize the overall discussion of the paper along with an explanation of some key challenges in Section~\ref{sec:challenges}. Finally, some concluding remarks are drawn in Section \ref{sec:conclusion}.

\section{Background on P2P Sharing}\label{sec:background} 
\subsection{Connected community}\label{sec:ConnectedCommunity}It is a well-known fact that coordinated DER brings several advantages to the energy system including reduced cost of transmission and distribution systems, reduced grid power losses, and a larger share of zero-carbon technologies~\cite{RalphSims_BookChp_2007}. However, to reap these benefits, prosumers have to play active roles in providing energy services~\cite{Peck_IEEESpectrum_Oct_2017} - a need that ultimately introduces the concept of a \emph{connected community}. 

A connected community enables collaboration between prosumers to take initiatives that results in collaborative solutions on a local basis to facilitate the development of sustainable energy technologies~\cite{ThomasBauwens_March_2016}. While the incumbent traditional energy grid suffers a lack of trust from the public, connected community enhances social acceptance of technology at the local level - steered by trustworthy individual prosumer and organizations rooted in the local community~\cite{ThomasBauwens_March_2016}. Essentially, a connected community consists of a group of efficient and interactive prosumers, such as owners of buildings with diverse and flexible end-user equipment with some electric vehicle (EV) charging infrastructure\footnote{https://www.energy.gov/eere/articles/department-energy-releases-request-information-potential-funding-grid-interactive.}, that can collectively work together to maximize the grid efficiency without compromising prosumers needs, comfort and convenience. Connected communities rely on smart technology, DER, flexible loads, and grid integration in order to reduce energy use and peak demand, improve energy efficiency, while at the same time, improve user's experience without compromising any privacy and security. 

\begin{figure}[t!]
\centering
\includegraphics[width=0.45\columnwidth]{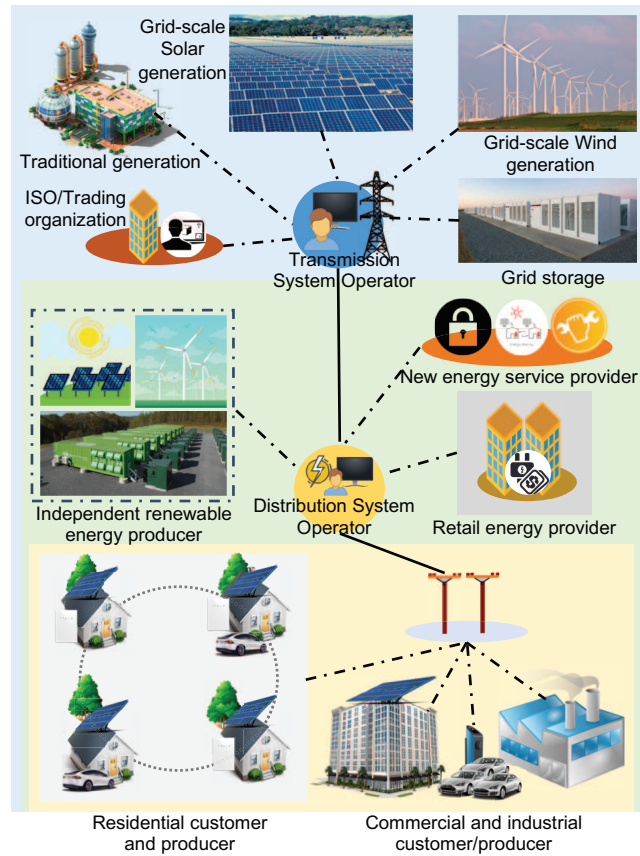}
\caption{This figure shows an overview of various elements of a connected community where selected participating prosumers such as households with renewable generators, community storage, and independent renewable energy providers generate electricity, which is shared among other energy users within the network. The energy service provider facilitates the platform for sharing within the network and provides additional services such as home energy management and energy donation services if necessary. Utility networks such as DSO is in charge of delivering electricity within the community. This figure inspired by~\cite{Abrishambaf_ESR_Nov_2019}.}
\label{Fig:Figure1}
\end{figure}

The physical electrical connection between different prosumers with a local community and across different communities is maintained by distribution and transmission lines of the power network - managed and maintained by distribution system operators (DSO) and the transmission system operator (TSO) respectively~\cite{Abrishambaf_ESR_Nov_2019}. The energy sharing services and relevant decision-making processes, on the other hand, can be provided by different application-specific energy service providers via transactive energy frameworks~\cite{Mohsen_Energies_Apr_2020} using digital communication, artificial intelligence, and signal processing techniques over the virtual network. An example of a connected community, in which energy is managed between different entities using a transactive energy framework is shown in Fig.~\ref{Fig:Figure1}. 

\subsection{P2P energy sharing systems}\label{sec:P2PsharingSystems}P2P energy sharing is a branch of transactive energy that considers prosumers' perspectives while at the same time ensuring that the system is operating safely and efficiently. In P2P sharing, prosumers can actively participate in the energy market, negotiate the price with other peers within the connected community, and then trade their energy and flexibility services as forms of either watt~\cite{Alam_AE_Mar_2019} or negawatt~\cite{Tushar_NE_2020}. With the power of setting the terms of sharing and delivering goods and services, it has been shown in several studies that the gain that prosumers can reap from participating in P2P sharing could be substantial~\cite{Tushar_TSG_Jan_2020}. Meanwhile, P2P sharing also benefits the grid in terms of reducing the peak demand~\cite{Tushar_TSG_Mar_2020}, reserve requirement~\cite{Andoni_RSER_Feb_2019}, operational cost~\cite{Esther_AE_Jan_2018}, and improving reliability~\cite{Thomas_Nature_2018}.

To facilitate this beneficial energy management scheme within connected communities, the P2P sharing system is divided into two layers - the physical layer and the virtual layer~\cite{Esther_AE_Jan_2018}. In the physical layer, the main elements are grid connection, smart metering, and communication infrastructure.
\begin{itemize}
\item\emph{Grid connection:}~It is important to define the connection points of the main grid for balancing the demand and generation of energy for both grid-connected and islanded microgrid based P2P sharing systems. The performance of the P2P sharing system can be monitored and evaluated by connecting smart meters in those connection points~\cite{Tushar_SPM_July_2018}.
\item\emph{Smart metering:}~In a P2P energy sharing system, each prosumer is equipped with a smart meter capable of deciding whether a prosumer should share its energy with other peers within the community based on the available information on demand, generation, and market condition. Smart meters also have the ability to communicate with each other through suitable communication protocols.
\item\emph{Communication infrastructure:}~A communication infrastructure is necessary within a P2P energy sharing system to discover prosumers and facilitate information exchange among them. The adopted communication infrastructure within a connected community needs to fulfill the requirements necessary for recommended system performance including latency, throughput, reliability, and security~\cite{Jogunola_Energies_Dec_2018}.
\end{itemize}

The virtual layer, on the other hand, comprises information systems, market operation, pricing, and energy management system.
\begin{itemize}
\item\emph{Information system:}~The information system helps prosumers within a P2P energy sharing system to decide on energy parameters by integrating them to a suitable market platform with equal access to each participant, monitor the market operation, and imposing constraints on prosumers' decisions, if required, for the network security and reliability purposes.
\item\emph{Market operation:}~The purpose of the market operation is to enable prosumers to experience an efficient energy sharing process by providing services to match the buy and sell orders in real-time. The different time horizon of the market operation enables participants to share their resources with different community members at various time slots of trading at a price that commensurate with the status of demand and supply of energy within the community. 
\item\emph{Pricing mechanism:}~The pricing of P2P sharing balances between the energy demand and supply within the connected community. Depending on the regulation within the region, energy prices may or may not include surcharges, taxes, and subscription fees. Regardless of types, all pricing schemes reflect the state of the energy within the connected community.
\item\emph{Energy management system:}~The energy management system (EMS) of a prosumer is responsible to bid in the market on behalf of the prosumer to share energy and simultaneously ensures the security of prosumer's energy supply. The decision of an EMS on the participation in energy sharing is triggered by real-time demand and supply information of the respective prosumer and the rules set by the prosumer on various market parameters including price, source of energy, and roles in the market (e.g., watt or negawatt sharer).
\end{itemize}

\begin{figure}[t!]
\centering
\includegraphics[width=0.45\columnwidth]{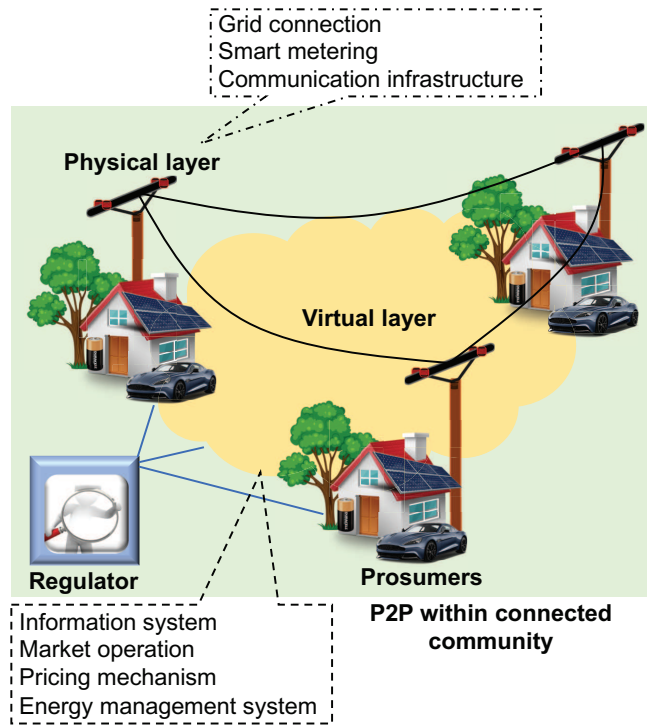}
\caption{Demonstration of different layers of P2P energy sharing systems within a connected community. The virtual layer provides a platform for the participating prosumers to decide on their energy trading parameters. The physical layer, on the other hand, facilitates the actual transfer of energy between the seller and the buyer once the decision is made in the virtual layer and payment is completed.}
\label{Fig:Figure2}
\end{figure}

In addition, a P2P energy sharing market has other two elements including prosumers and regulators. Clearly, a sufficient presence of prosumers is vital for the success of P2P energy sharing within connected communities whereas the regulation within a region decides whether the P2P energy sharing system should be facilitated within existing energy market and supply systems.  A demonstration of different layers of the P2P energy sharing system and relevant elements is shown in Fig.~\ref{Fig:Figure2}.
\subsection{P2P energy sharing market}\label{sec:P2PSharingMarket}
\begin{figure*}[t]
  \centering
  \begin{minipage}[c]{0.5\textwidth}
    \centering
    \includegraphics[scale=0.45]{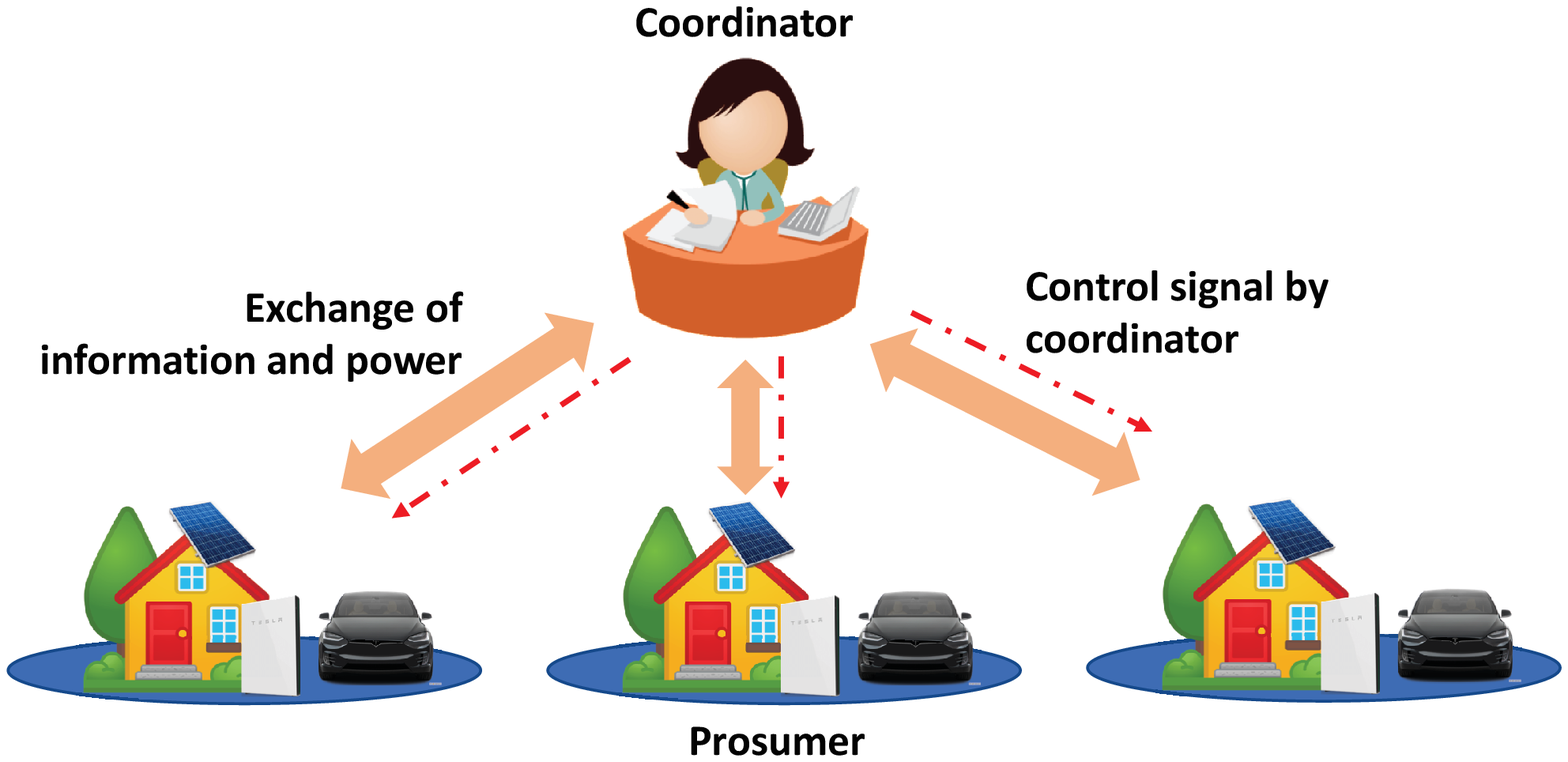}
    \subcaption{Coordinated market.}
    \label{fig:Coordinated}
  \end{minipage}
~
  \begin{minipage}[c]{0.5\textwidth}
    \centering
    \includegraphics[scale=0.45]{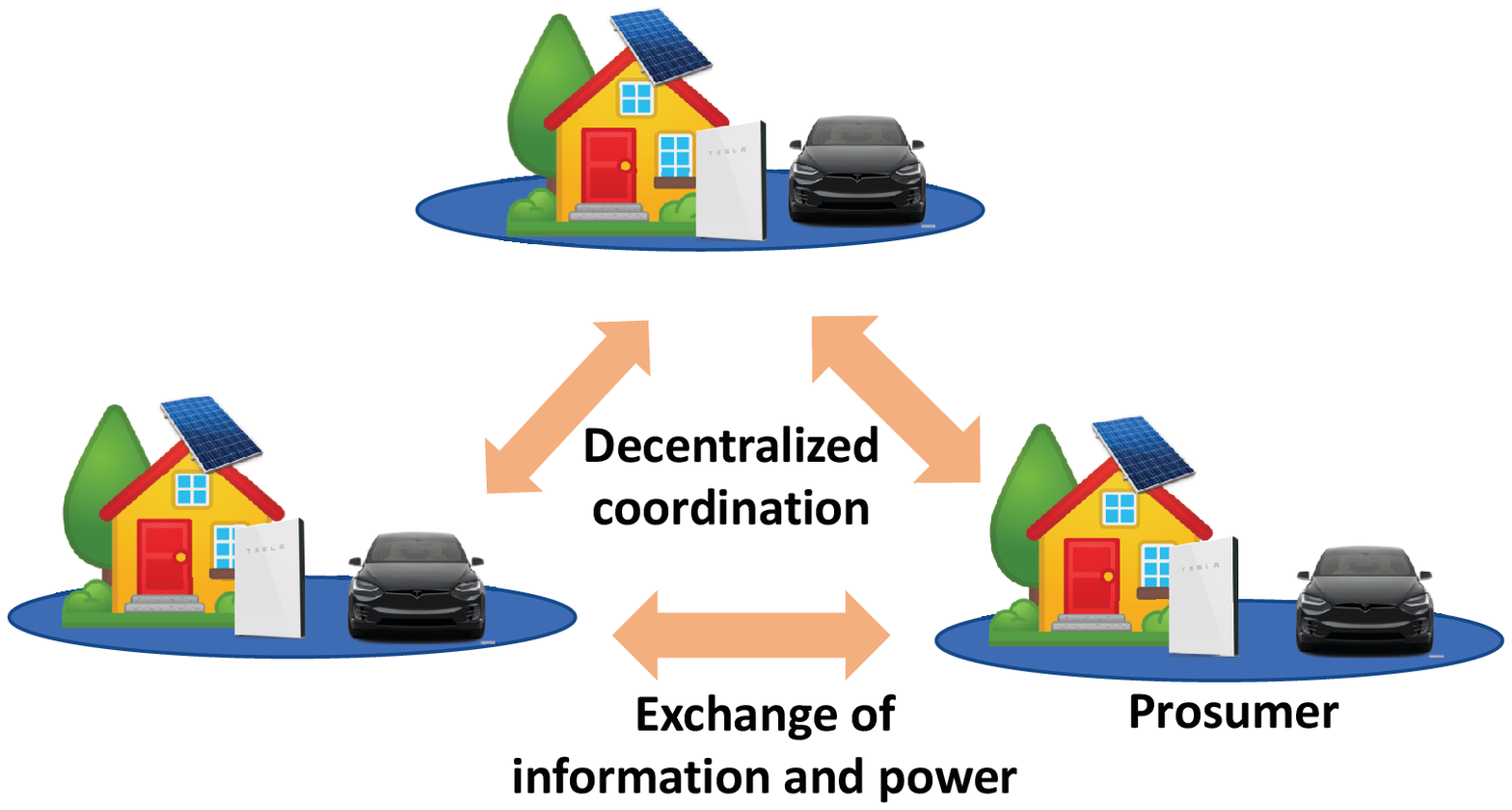}
    \subcaption{Decentralized market.}
    \label{fig:Decentralized}
  \end{minipage}
\\
  \begin{minipage}[c]{ \textwidth}
    \centering
    \includegraphics[scale=0.5]{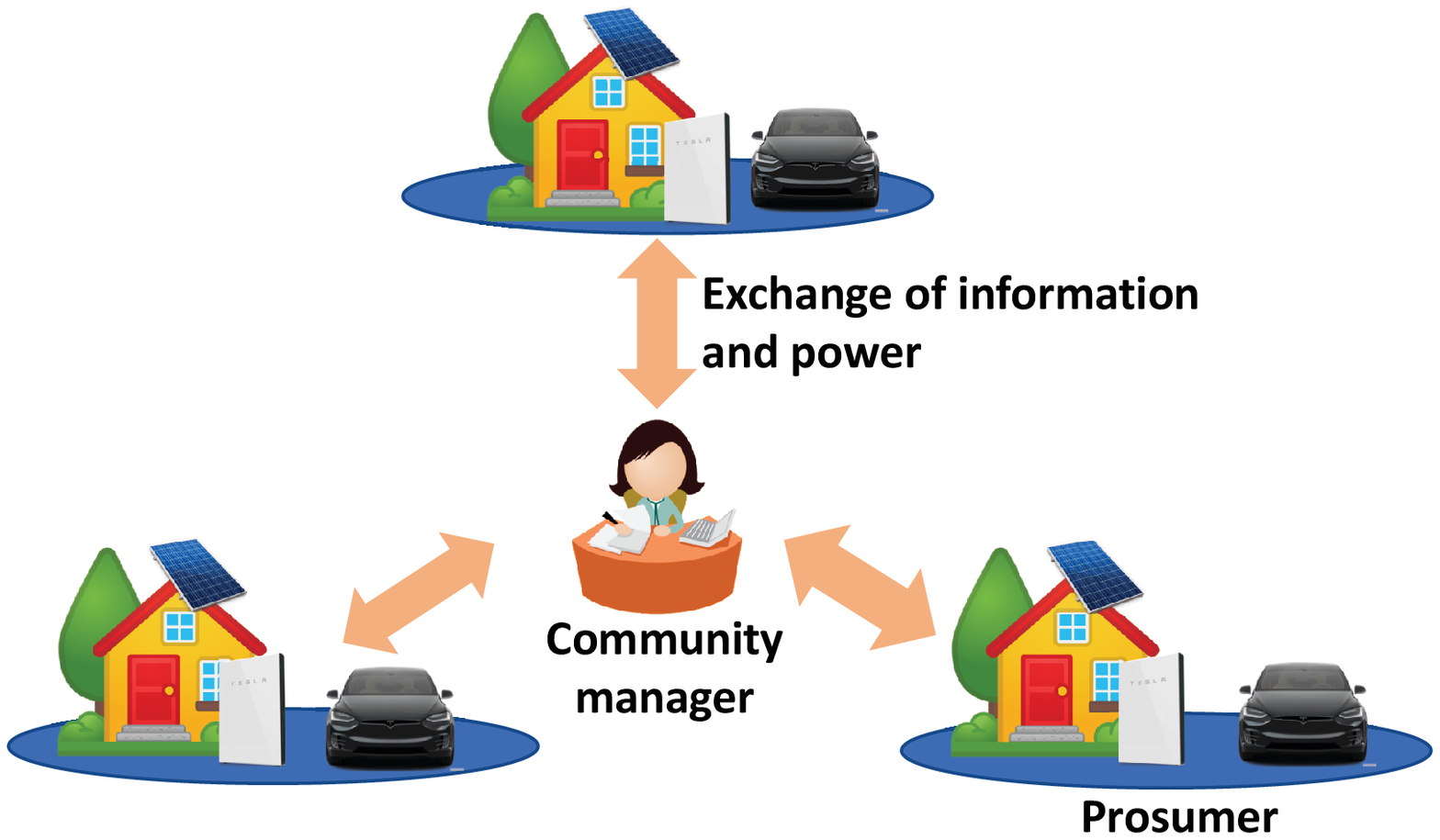}
    \subcaption{Community market.}
    \label{fig:Community}
  \end{minipage}
  \caption{Based on the coordination process of sharing and information communication, this figure illustrates different types of market that facilitate P2P sharing within an energy network. In a coordinated market, both sharing and communication are done in a centralized fashion whereas in a decentralized market both are performed through decentralized method. In a community market, the communication process is centralized, but the sharing process is decentralized.}
  \label{fig:P2PMarkets}
\end{figure*}

Considering how the trading process is performed and how the communication of information takes place among the participants, the P2P energy market can be divided into three categories: coordinated market, decentralized market, and community market, as shown in Fig.~\ref{fig:P2PMarkets}.
\subsubsection{Coordinated market}In a coordinated market, both the trading process and the communication of information are done in a centralized fashion (Fig.~\ref{fig:Coordinated}). That is, a centralized coordinator communicates with each peer within the network and directly control the export and import limit of energy that the prosumers share among themselves through P2P trading. Once the sharing is complete, the revenue of the entire connected community is distributed among the prosumers by the coordinator according to pre-set rules. An example of such a pre-set rule is proportional allocation~\cite{Tushar_TSG_May_2016}.  While each peer does not communicate and negotiate energy sharing parameters with other peers in a coordinated market, they can influence the decision of energy sharing parameters by independently decide their energy and price before sharing that information with the coordinator. A key advantage of the coordinated market is the outcome that maximizes social welfare~\cite{Zhou_Engineering_June_2020}. However, with increasing penetration of DER, the computational burden of P2P sharing could be extensive~\cite{Papadaskalopoulos_TPWRS_Nov_2013}. Further, due to direct control of peers' flexible loads and DER, a coordinated market can potentially compromise the privacy of the prosumers. More discussion on coordinated markets can be found in \cite{Luth_AE_Nov_2018} and \cite{Hou_TII_June_2019}.
\subsubsection{Decentralized market}In a decentralized market (Fig.~\ref{fig:Decentralized}), peers can directly communicate with one another within the connected community and decide on their energy trading parameters without the involvement of any centralized coordinator~\cite{Tushar_AE_June_2019}. Thus, in a decentralized market, both the trading process and the communication of information are done in a decentralized fashion. The main advantage of a decentralized market is that prosumers are in full control of their decision-making process, e.g., they can easily decide whether to participate in energy sharing or not at any given time slot and their privacy is well protected~\cite{Tushar_TSG_Jan_2020}. Further, the scalability of the decentralized market is also exceptional~\cite{Zhou_Engineering_June_2020}. Due to prosumer-centric properties, decentralized markets serve the prosumers better than the coordinated market. 

However, due to a lack of centralized control~\cite{Thomas_TSG_Mar_2019}, the efficiency of decentralized markets is relatively low, and social welfare does not attain the maximum value. As the overall energy that can be traded within the community is not very clear to third parties such as network operators, retailers, and transmission system operators, managing the decentralized market is more difficult for service providers due to the challenge of maintaining network constraints and  improving the operational efficiency of the power system. To maintain such a market, sometimes network operators need to take drastic measures such as load curtailment and blocking peers from the network~\cite{Archie_P2P_Sept_2019} in order to maintain the reliability of the grid. Further examples of P2P sharing in the decentralized market can be found in \cite{Sorin_TPWRS_Mar_2019} and \cite{Khorasany_TII_July_2020}.
\subsubsection{Community market}In a community market (Fig.~\ref{fig:Community}), the trading process of decentralized although the communication between the participating prosumers is done in a centralized fashion. In this market, a community manager acts as a coordinator of P2P energy sharing among the prosumers. However, unlike the centralized market, the community manager cannot directly control the export and import of energy by different prosumers within the market. Rather, the community manager influences the prosumers to participate in P2P sharing indirectly via suitable pricing signals~\cite{Tushar_TSG_May_2016}. Thus, in a community market, prosumers need to share limited information with the community manager while, simultaneously, they can maintain a higher level of privacy~\cite{Sousa_RSER_Apr_2019}. Further, through indirect control, the autonomy of prosumers in deciding their energy trading parameters is also preserved. One core focus of the community market-based energy literature is to design suitable pricing schemes that can facilitate P2P sharing  and at the same time can provide energy services to different entities within the network. Pricing schemes also focus on engaging a large number of prosumers in energy sharing.  Different energy sharing mechanisms within community markets have been discussed in \cite{Paudel_TIE_Aug_2019}, \cite{Gonzalez_IEM_Dec_2018}, and \cite{Moret_TPWRS_EA_2018}. 

A summary of different types of market for P2P sharing is shown in Table~\ref{table.markets}. We further note that in a connected community, it is also possible that different types of markets coexist at the same time. For example, both decentralized and community market can exist under a composite market paradigm that possesses the characteristics and advantages/disadvantages of both markets. For details on P2P energy sharing markets, please see \cite{Sousa_RSER_Apr_2019} and \cite{Zhou_Engineering_June_2020}.
\begin{table*}[t]
\centering
\caption{Summary of different type of markets for P2P energy sharing. The market structure is based on the characteristic of trading process and the method of communication of information among participants in the market.}
\begin{tabular}{c c c c c c }
\hline
\multirow{2}{*}{\bfseries Type of market}&\multicolumn{2}{c}{\bfseries Trading process} & \multicolumn{2}{c}{\bfseries Communication of information}  & \multirow{2}{*}{\textbf{References}}\\
& Centralized & Decentralized & Centralized & Decentralized&\\
\cline{1-6}\\
Coordinated market & \checkmark & & \checkmark & & \cite{Tushar_TSG_May_2016,Zhou_Engineering_June_2020,Papadaskalopoulos_TPWRS_Nov_2013,Luth_AE_Nov_2018,Hou_TII_June_2019}\\\\
Decentralized market & & \checkmark & & \checkmark & \cite{Tushar_AE_June_2019,Tushar_TSG_Jan_2020,Thomas_TSG_Mar_2019,Archie_P2P_Sept_2019,Sorin_TPWRS_Mar_2019,Khorasany_TII_July_2020}\\\\
Community market & & \checkmark & \checkmark & & \cite{Tushar_TSG_May_2016,Sousa_RSER_Apr_2019,Paudel_TIE_Aug_2019,Gonzalez_IEM_Dec_2018,Moret_TPWRS_EA_2018}\\\\
\cline{1-6}
\end{tabular}
\label{table.markets}
\end{table*}
\subsection{Technologies behind P2P sharing}\label{sec:TechnologiesP2PSharing}Successful establishment of P2P sharing of energy has been possible due to innovations and developments of a number of technologies. What follows is an overview of technologies that have enable sharing of energy between prosumer within connected communities.
\subsubsection{Distributed ledger}\label{sec:Ledger}Due to the nature of the P2P energy sharing mechanism, data security, data privacy, data integrity, and speed of financial transactions between prosumers become very critical~\cite{Siano_Systems_Sept_2019}. As such, distributed ledger technology has been proven to be very effective in addressing these concerns by providing prosumers with a platform with transaction security to exchange information among themselves for both energy and economic transactions without resilience on trusted third parties. Thus, it is necessary when the communication of information is decentralized, i.e., decentralized market. The main parts of any distributed ledger technologies include ledgers, smart contracts, and consensus protocols~\cite{Zia_Access_Jan_2020}. Ledgers record key information and data about the participants whereas smart contracts define participants' preferences to ensure the implementation of agreed terms between two or more parties. The objective of consensus protocols is to validate transactions. Some examples of distributed ledger technology include Blockchain~\cite{Li_TII_Aug_2018}, Hashgraph~\cite{Hashgraph_TVT_2020}, Holochain~\cite{Holochain_2020}, Directed acyclic graph~\cite{DAG_AppliedEnergy_2015}, Hyperledger~\cite{Hyperledger_IBM_2019}, Ethereum~\cite{Ethereum_Access_2019}, and Algorand~\cite{Chen_TCS_July_2019}. A detailed survey of different distributed ledger technology can be found in \cite{Zia_Access_Jan_2020}.
\subsubsection{Internet-of-Things}\label{sec:IOT}Internet-of-Things (IoT) is a platform that enables devices to communicate with one another and with humans over the internet to achieve various objectives, such as energy saving, condition monitoring, predictive maintenance, and remote monitoring and control~\cite{Tushar_SPM_Sept_2018}. An important part of P2P energy sharing is that prosumers' monitor their own energy generation and demand~\cite{Ali_TCE_Nov_2017}, control and schedule the energy consumption pattern of various appliances~\cite{Nasim_AE_Aug_2018}, and set the rules for the devices~\cite{Ethan_AE_Apr_2019} to participate in the energy sharing over the electricity network based on the information available in the energy sharing platform. IoT has made all these tasks possible and thus contributes extensively to prepare P2P sharing as an energy management technique in the local electricity market. Applications of IoT have been extensively discussed in \cite{Bedi_IOT_Apr_2018}, and \cite{Reka_RSER_Aug_2018}. 
\subsubsection{Artificial intelligence}\label{sec:AI}Artificial intelligence (AI) is a modern breakthrough in computational techniques that has the potential to revolutionise automation and decision-making of agent-based systems. For example, AI could be a vital part of the `smart' of smart grid energy sharing~\cite{Sarvapali_ACM_Apr_2012}. In general, AI refers to computational techniques that simulate human intelligence in machines to enable them to think and act like humans. Machines with AI exhibit characteristics associated with human mind such as learning and problem-solving~\cite{Yuzchen_AE_Feb_2018}. For P2P sharing, it is important to learn the energy usage pattern of different flexible loads within a building as well as understand prosumers' responses in terms of bidding energy amount and price in various market conditions. AI has been shown to be very effective to capture these learning objectives through reinforcement learning~\cite{Jose_AE_Feb_2019}, deep learning~\cite{Konstantakopoulos_AE_Mar_2019}, and artificial neural network~\cite{Reynold_AE_May_2018}.
\subsubsection{Responsive buildings}\label{sec:GEB}Responsive buildings, also known as grid-efficient buildings~\cite{Perry_GEB_Oct_2019}, refer to buildings that have capabilities to respond to incentive signals sent from the grid, other buildings, or third parties by altering their energy generation, consumption, and sharing behavior. Therefore, responsive buildings have the capability to monitor and control their real-time energy generation and dispatch, optimize the energy usage behavior according to occupant needs and preferences, provide energy services to the grid and other energy entities within the connected community. To exhibit these capabilities, responsive buildings are equipped with 1) reliable and low-latency two-way communication facilities to communicate with devices, appliances, and other responsive buildings within the community; 2) intelligence management system to monitor, predict, and learn from occupants' behavior, weather forecast, and market condition, and subsequently take complex and intelligent actions that adapt dynamically over multiple time slots and various conditions; and 3) a secured and trusted platform that is resilient against cyber attack from unauthorized sources. An overview of responsive buildings can be found in \cite{Neukomm_GEB_Apr_2019}.
\subsubsection{Controllable DER} DERs without coordination and control introduce reverse power flows, voltage rise, and increased fault currents within the electricity network~\cite{Arturo_PhDThesis_2009} that can lead to a widespread catastrophic event like a blackout~\cite{Yan_TPWRS_Sept_2018} and therefore trigger the need for network equipment reinforcement. To avoid this, significant efforts have been reported recently in advancing the techniques for controlling DERs like PV inverters, battery storage, electric vehicles, and demand response asset (flexible loads). Now, PV inverters are smart enough to self-adjust their active and reactive power injection to the electricity network in response to the network condition~\cite{Weckx_TSE_Oct_2016}, whereas cutting edge algorithms and techniques are available to opportunistically charge and discharge the battery storage to provide network services~\cite{Jahedul_RSER_Oct_2020} and reduce cost~\cite{Li_TSG_July_2019} by the prosumers. Coordinated control of electric vehicle charging/discharging along with the intelligent management of flexible loads now enables prosumers to enjoy the benefit of having mobile storage~\cite{Kikusato_TSG_May_2019} and participate in P2P trading using the energy stored in the battery of their vehicles~\cite{Kang_TII_Dec_2017}. Further, these DERs represent new points of control on distribution feeders that can be aggregated meaningfully to impact the bulk electricity system \cite{Arnold_TPWRS_Jan_2018} via P2P trading.
\subsubsection{Design innovation}\label{DI}Design innovation (DI) is a human-centred and interdisciplinary approach that integrates technology with user experience and positively reflects users' preferences in energy management schemes such as P2P sharing. As P2P has been identified as a socio-technical energy management scheme~\cite{Kristina_Thesis_2012}, rather than just a technical scheme, taking participants' preferences into account while designing P2P mechanism is of paramount importance~\cite{Tushar_AE_June_2019}. As such, DI, through its 4D model (discover, define, develop and deliver), provides an excellent set of tools that can be applied for managing energy generation and consumption within buildings in a prosumer-centric way~\cite{Tushar_Energy_Apr_2020} with a view to enable the buildings to participate in the energy sharing market. 
\subsubsection{High speed communication}\label{sec:Communication}The term \emph{connected community} clearly emphasizes the need for suitable communication infrastructures for P2P energy sharing with the capability to operate remotely and interact with various devices within the connected community with low-latency. Recent advancement in high-speed communication such as fifth-generation communication (5G) can fulfill these requirements with its wide range of network capabilities and abilities to support highly demanding services.  Examples of supports that 5G can provide to advance P2P energy sharing include, but are not limited to, enabling communication between massively interconnected devices, enable operations and manipulation of physical objects over distance with reliability, and ensure low latency response~\cite{Sachs_ProceedingIEEE_Feb_2019}.
\begin{figure}[t]
\centering
\includegraphics[width=0.75\columnwidth]{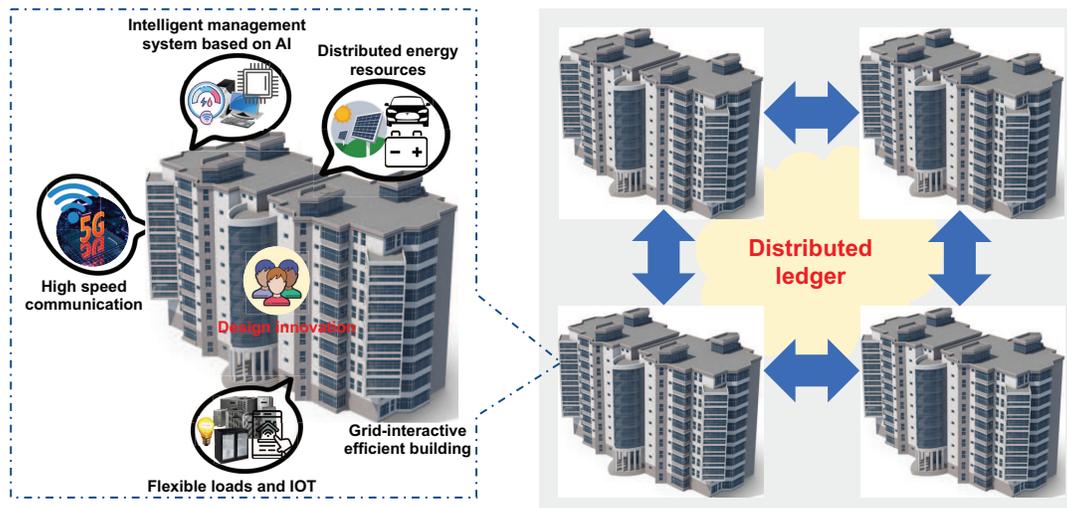}
\caption{A graphical overview of how technologies enable prosumers to share their energy within a connected community. Distributed ledger facilitates the information exchange and transaction among the participating entities of energy sharing. Meanwhile, artificial intelligence, IoT, and controllable DER critical for producing energy, e.g., by intelligently managing energy within the households and setting suitable price of sharing,  within the community to be shared by its elements. High speed communication ensure low latency energy transaction whereas design thinking establishes prosumer-centric outcome of overall energy sharing within the community.}
\label{Fig:OverviewTechnology}
\end{figure}

A graphical overview of how different technologies enable prosumers to share their energy within a connected community is shown in Fig.~\ref{Fig:OverviewTechnology}.
\section{Advancement of P2P in Various Domains}\label{sec:AdvancementofP2P} 
\begin{figure}[t]
\centering
\includegraphics[width=0.75\columnwidth]{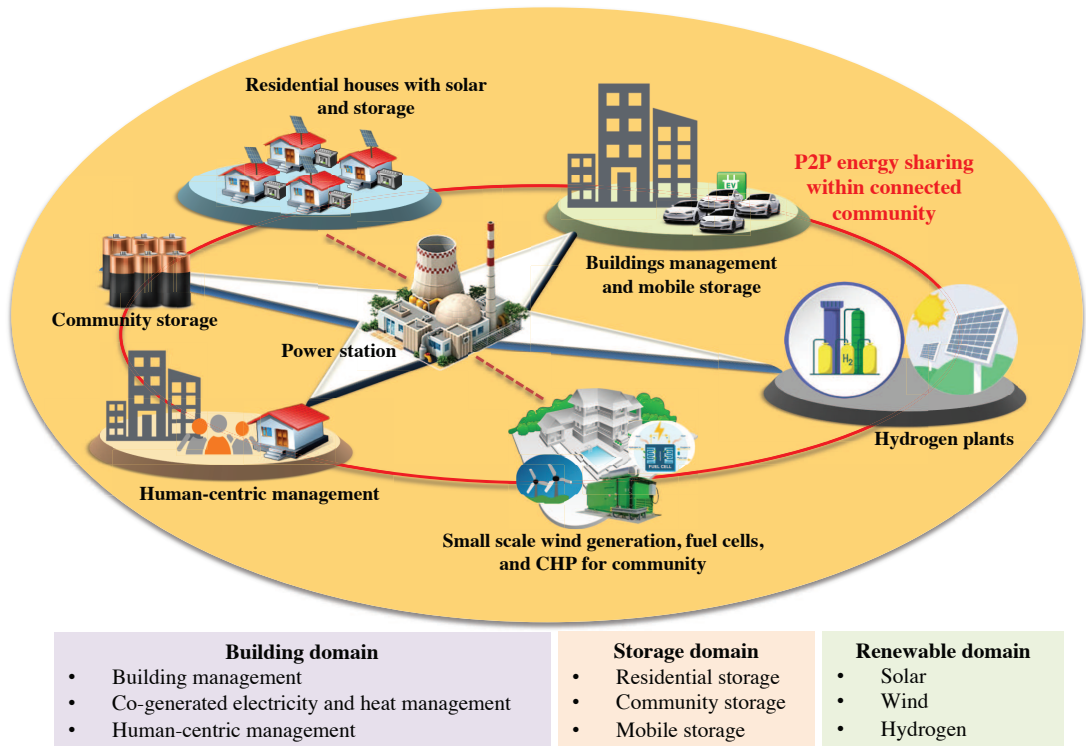}
\caption{A graphical presentation of different domains in which significant advancements have been reported in the literature over the past few years. In the building domain, the main focus has been on building management, co-generated electricity and heat management, and human-centric management. In the storage domain, residential storage domain dominates the literature. However, community storage and mobile storage have also received some attention in recent times. In the renewable domain, solar, wind, and hydrogen have been the foci of discussion.}
\label{Fig:Domains}
\end{figure}

In the last decade, advancements in technological research that are directly or indirectly contributing to the successful sharing of energy in the P2P network have been extensive. Considering which network elements have mainly been utilized to either directly conduct P2P sharing or leverage the sharing of energy, existing studies in the literature can be divided into three domains: building domain, storage domain, and renewable domain. Each domain has its own distinct characteristic and conditions for participating in P2P sharing. For example, what kind of prosumer in each domain can participate in P2P sharing and what are the conditions they need to conform to implement P2P sharing within the domain. Thus, the proposed domain-specific discussion would help the reader to understand the challenges and conditions that each prosumer of the domain needs to address for participating in P2P sharing. A graphical demonstration of these different domains in the energy network is shown in Fig.~\ref{Fig:Domains}.
\subsection{Building domain}\label{sec:BuildingDomain}To participate in P2P energy sharing in this domain, at least some buildings within the connected community should have enough provisions to produce energy to be shared by the community members. For example, a building can be equipped with DER such as storage and rooftop solar from which it can use the generated energy to participate in P2P sharing. However, having storage and renewable generator is not mandatory for buildings to produce energy. A building may use other DERs such as flexible loads, interruptible appliances such as HVAC and lights, and electric vehicles to control its energy consumption and thus reduce demand to share either watt or negawatt with the participants. However, a key condition that each building needs to fulfill before utilizing the flexibility for energy sharing is customer preference and comfort. In other words, energy management within the building should be human-centric to confirm that human convenience and preferences are prioritized. Further, as the use of fuel cells and combined heat and power is becoming increasingly popular in the building, the P2P sharing platform should have enough adaptability to accommodate this emerging energy system. Given this context, existing studies in the building domain have focused on P2P sharing from three different points of view: 1) building management, 2) co-generated electricity and heat management, and 3) human-centric management.

\subsubsection{Building management}\label{sec:BMS}To enable buildings to participate in P2P energy sharing with one another, it is critical that buildings can produce surplus through proper management and scheduling of their flexible loads including heating, ventilating, and air-conditioning (HVAC), lighting, and other adjustable loads. Consequently, in the building domain, we focus on how the building management system (BMS)~\cite{Manic_IEM_Mar_2016} within buildings can create energy surplus or reduce energy deficiency in order to enable the building to participate in P2P energy sharing~\cite{RuiJing_AE_Mar_2020}.  For example, the authors in \cite{Alam_AE_Mar_2019} and \cite{Alam_EE_Dec_2017} show how smart homes can control HVAC and other flexible loads in order to participate in P2P energy sharing with other smart homes within the connected community. Now, based on building types, BMS finds its application mainly in two kinds of buildings: residential building~\cite{Ippolito_ENB_Feb_2014} and commercial buildings~\cite{Lazos_RSER_Nov_2014}. In both types of buildings, energy management is done through the management of HVAC control~\cite{WenTai_TETC_Sept_2019}, lighting control~\cite{Marco_AE_June_2020}, and scheduling of flexible loads~\cite{PDu_TSG_June_2011}.
\paragraph{HVAC control}HVACs are one of the major contributors to energy demand and, at the same time, have great potential to help save energy and provide energy services~\cite{Raja_TETCI_2020} such as P2P sharing. To manage HVAC systems within buildings, a number of strategies have been reported in the literature including controlling the set-point temperature and limiting the air distribution and cooling system of the HVAC system.

\emph{Set-point temperature control:}~The set-point temperature control of the HVAC system within buildings is motivated by ASHRAE 55-1992~\cite{ASHARE_2006} guidelines and is done by adopting short-term curtailment~\cite{Gu_Ashrae_2012}, departure and arrival preparation~\cite{Yang_ENB_Aug_2014}, pre-cooling~\cite{Su_IFAC_2014}, modified pre-cooling~\cite{Gayeski_HVAC_Sept_2012}, programmable thermostat adjustment~\cite{Nikdel_BNE_Feb_2018}, and temperature reset~\cite{WenTai_TETC_Sept_2019} approaches. Further, artificial intelligence-based control schemes such as in~\cite{Ngarambe_ENB_Mar_2020}, \cite{Lork_AE_Oct_2020}, \cite{Zahra_EPSR_Nov_2020}, and \cite{Valladares_BNE_May_2019} have also been used extensively for the set-point temperature control of HVAC systems.

\emph{Systematic adjustment:}~The systematic adjustment of HVAC systems is achieved by placing limits on air distribution and cooling system equipment in the HVAC system~\cite{Raja_TETCI_2020}. Although it is difficult to re-balance the system, e.g. due to the imbalance of cooling (chilled water and supply air) distribution to individual zones of a building, via systematic adjustment, it provides a fast power reduction performance.
\paragraph{Lighting control}Lighting is the third-largest electricity consumer in both commercial and residential buildings~\cite{Nagy_ENB_May_2015}. Thus, like the HVAC, lighting control also has the potential to create enough surplus or demand flexibility for buildings to trade in the local energy market. With the advancement of IoT and sensor technology, a number of energy savings techniques have been introduced for building lighting control. 

\emph{Occupancy-based strategy:}~With proper occupancy-based lighting control, the energy usage for lighting can be reduced by between 20\% and 60\% compared to the case without control~\cite{Bakker_BNE_Feb_2017}. As such, a large number of studies have occupancy as the key decision-making parameter to decide the lighting status of a building. The authors in \cite{Oldewurtel_AE_Jan_2013} utilize an integrated room automation technique to control the lighting of an office building using the occupancy information in the control scheme. In \cite{Park_BNE_Jan_2019}, LightLearn - a reinforcement learning-based lighting control scheme - is developed using occupant preferences that successfully balance occupant comfort and energy consumption. Similar examples of occupancy-based lighting control within buildings can also be found in \cite{Zou_ENB_Jan_2018} and \cite{Labeodan_ENB_Sept_2016}.

\emph{Sunlight-based strategy:}~For buildings receiving sunlight, lighting control schemes based on the availability of sunlight can provide the maximum amount of savings. In \cite{Meugheuvel_ENB_June_2014}, the authors propose a distributed lighting control mechanism that takes daylight and occupancy status of the building into account, whereas \cite{Liu_ENB_Sept_2016} has developed a fuzzy logic controller for saving energy in smart LED lighting systems considering sunlight within the space. Further studies that have focused on lighting control in buildings can be found in \cite{Haq_RSER_May_2014}.

\paragraph{Flexible load schedule and control}The term \emph{flexible load} refers to the load within buildings that is controllable and can be scheduled to operate at different times of the day and night, and therefore is a key factor behind the ability of a building to participate in P2P sharing. Examples of flexible load include washing machines, dishwashers, hot water pumps, and electric vehicles (EV). Flexible loads are required for a building to participate in demand response, for example, to participate in the P2P market~\cite{Alam_EE_Dec_2017} and to provide demand flexibility to the grid~\cite{Li_CSEE_Dec_2018}. The scheduling and control of flexible loads are usually done either via direct~\cite{Georges_AE_Feb_2017} or indirect control~\cite{NaveedHassan_TSG_Nov_2015}. Direct control schemes are mainly incentive-based programs, in which there is an agreement between the customers and the service provider that provides the program coordinator with some degree of access to directly schedule, reduce or disconnect flexible loads within buildings, and building owners receives some incentives in return for agreeing to provide the coordinator with access to the building.  Examples of direct control load programs include interruptible load control~\cite{Huang_TPWRS_Aug_2004}, interruptible tariffs~\cite{Bhattacharya_TPWRS_May_2000}, demand-bidding program and emergency program~\cite{Habib_TEES_Nov_2016}.

Indirect control schemes, on the other hand, rely on different pricing, where customers are offered time-varying rates that reflect the value and cost of electricity at different times during the day~\cite{Wang_TSG_July_2020} and are thus encouraged to individually manage their loads either by reducing their consumption or shifting their energy activities from peak periods to off-peak hours~\cite{Andersen_AE_Dec_2017} to create enough flexibility to participate in P2P sharing \cite{Tushar_TSG_Mar_2020,Guerrero_RSER_Oct_2020}. A collection of research papers that have investigated this particular issue of load scheduling and control can be found in the review article~\cite{Hussain_EJ_June_2018}. 

\begin{figure}[t]
\centering
\includegraphics[width=0.6\columnwidth]{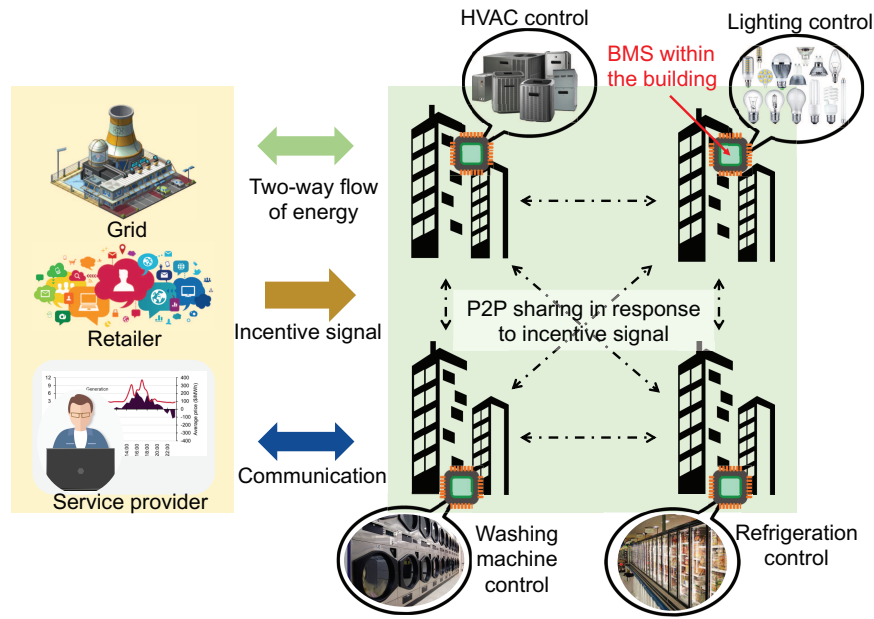}
\caption{A demonstration of how buildings can be incentivized by the grid and energy service providers to participate in P2P sharing by managing their flexible loads.  For example, in response to an incentive signal from the retailer, a building can control its flexible loads, HVAC, and lighting systems to produce some surplus energy to share with other buildings within the connected community. This figure is inspired from the scheme proposed in \cite{Tushar_TSG_Mar_2020}.}
\label{Fig:BuildingDomain}
\end{figure}

A demonstration of how different energy savings can enable buildings to participate in P2P sharing is demonstrated in Fig.~\ref{Fig:BuildingDomain}.
\subsubsection{Co-generated electricity and heat management}\label{sec:water}Despite relatively high system costs, the use of fuel cells and combined heat and power (FC-CHP) is becoming popular in residential housing due to high thermal and electricity efficiency, low emissions, and reduced electricity bills. As a result, FC-CHP co-generation system is becoming another popular DER for local generation and consumption. As P2P sharing has been established as a very promising technique for residents in connected communities to reduce their cost and improving system performance in terms of balancing local supply and demand, households with FC-CHP systems are also expected to share their co-generated heat and electric energy with neighboring households.  For example, in \cite{Wakui_Energy_Feb_2010}, the authors propose a mixed-integer linear programming approach to determine an operational strategy for a power interchange using multiple residential solid oxide fuel cell cogeneration systems for saving energy. An energy management system is developed in \cite{Aki_Hydrogen_Nov_2016} that achieves optimal operation of FC-CHP systems with energy exchange between households within a connected community.  An optimal strategy for managing multiple heat sources in a connected community is proposed in \cite{Aki_Energy_June_2018}. Finally, a distributed energy management system is developed in \cite{Tran_ECM_Oct_2018} that optimally schedules multiple combined heat and power systems in a connected community. 
\subsubsection{Human-centric management}\label{sec:HumanCentricManagement}Human-centric or prosumer-centric~\cite{Wayes_AE_Mar_2020} management can be defined as energy management schemes that consider human convenience and preferences as a priority and produce outcomes, which ultimately benefit the prosumers. Within the building domain, human-centric management schemes are closely aligned with HVAC control, lighting control, and flexible load management and control with a significant focus on human behavior and convenience. For example, a number of human-centric building management schemes have been studied in \cite{Jazizadeh_AE_June_2018,Jung_BE_July_2019,Adhikari_TSG_May_2020} and \cite{Jiang_EPSR_Sept_2020}. In \cite{Jazizadeh_AE_June_2018}, the authors propose a framework of a ubiquitous thermal comfort assessment for energy-efficient HVAC using RGB video images of human thermoregulation states. The same authors also study a comparative assessment of HVAC control strategies using personal thermal comfort and sensitivity models in \cite{Jung_BE_July_2019}. In \cite{Adhikari_TSG_May_2020}, a set of algorithms is proposed for controlling HVACs of a group of residential houses that a demand response aggregator can use to sell regulation service in the wholesale market consiering comfort requirements of the households.

In addition to human-centric appliance control within buildings for demand response that subsequently produces resources for trading in the P2P market, a number of studies are also proposed in recent times with human-centric outcomes that influence building owners/occupants to participate in the P2P sharing. For example, in \cite{Thomas_Nature_2018}, the authors show how P2P trading can be exploited to incentivize prosumers to coordinate with one another to produce a federated power plant in the smart grid. The authors in \cite{Tushar_AE_June_2019}, \cite{Tushar_Energy_Apr_2020}, and \cite{Tushar_Access_Oct_2018} utilize motivational psychology to demonstrate the effectiveness of p2p trading in motivating building occupants to put their renewable resources in the market for sharing. More examples of similar techniques of human-centric P2P sharing schemes can be found in \cite{Paudel_TIE_Aug_2019} and \cite{Wilkinson_ERSC_Aug_2020}.
\subsection{Storage domain}\label{sec:StorageDomain}In this domain, prosumers rely on their storage to participate in P2P energy sharing. A challenging aspect of P2P sharing in this domain is the management of storages of different size and ownership. For example, residential users have small-scale storage, whereas, at the community level, medium or large-scale storage is in use. With the emergence of electric vehicles, charging and discharging of a large volume of mobile storage devices also have an impact on the decision-making process of the prosumers. So prosumers should have access to an intelligent battery storage management algorithm that can enable prosumers to decide on how to prioritize between different type of storage within the connected community and condition upon the energy price, demand, and available energy in the storage and from other sources - rooftop solar, for example - coordinate the charging and discharging of battery for P2P sharing to maximize the overall benefit for the community. Another challenge is the space that is required for the installment of storage, in particular, for community storage. It might not always feasible to install storage within connected community for space and cost constraints. Nevertheless, with storage, prosumers can benefit from its opportunistic utilization in the P2P market via enjoying better return of investment by selling the energy at peak pricing period. They can take advantage of their battery to become completely off-grid, if needed, in case of emergency such as hurricane or bushfire. As such, to demonstrate how existing studies have captured these conditions and benefits of using storage in the P2P market, we have classified the research in the storage domain into three sub-domains: residential energy storage, community (shared) storage, and mobile storage.

\subsubsection{Residential energy storage}\label{sec:ResidentialStorage}The majority of research in P2P energy sharing focuses on the participation of residential prosumers in the market. Consequently, in the storage domain, most of the existing studies captured how various residential prosumers can utilize their storage devices in the P2P market. Considering who is the beneficiary of P2P sharing, the existing research in the residential storage domain can be divided into two categories. The first category of studies focuses on the benefit to the prosumers whereas the second kind of studies demonstrate how the grid can be benefitted as an outcome of the sharing.

\paragraph{Prosumers benefit}By participating in P2P sharing, a prosumer can reap economic benefits. For example, in \cite{Nguyen_AE_Oct_2018}, it is shown that a household can increase its savings by up to 28\% when it involves both its PV and storage in energy sharing. As a consequence, a large number of studies including recent articles like \cite{Long_AE_Sep_2018,Jan_EB_Feb_2019,Nizami_AE_Mar_2020}, and \cite{Shantonu_AE_Nov_2018} have shown how prosumers can engage their residential storage devices in P2P sharing. In \cite{Long_AE_Sep_2018}, the authors propose a two-stage control of prosumers PV-battery systems in a community microgrid to demonstrate an overall savings of the community of 30\%. To engage prosumers P2P sharing, a framework is proposed in \cite{Jan_EB_Feb_2019}, which shows that combined P2P trading and battery storage can lead to savings of almost $60\%$ compared to the case without P2P trading. Another mechanism is proposed in \cite{Nizami_AE_Mar_2020} to enable prosumers to participate in day-ahead P2P energy trading, which uses a bi-level optimization-based bidding strategy. The concept of energy loan is introduced in \cite{Shantonu_AE_Nov_2018} that can increase the system efficiency and social welfare of P2P sharing. 

While the use of residential storage can benefit the prosumers significantly, battery storage is also very expensive. The authors of \cite{Hangyue_IJEPES_July_2019} compare the costs of standalone battery systems versus the potential energy uses that could be curtailed to avoid those costs when going off grid, and find that in both cases these costs are significant, especially compared to striking P2P energy sharing agreements. In addition, extensive charging and discharging could damage battery life~\cite{Markus_IJER_Feb_2020}. Therefore, it is important for prosumers to decide whether to invest in their own storage or share community storage. Further, they also need to determine the most effective way for their battery storage devices to participate in P2P sharing. In \cite{Rodrigues_AE_Mar_2020}, the authors show that while owning a battery could be economically beneficial, community storage offers the prosumers an opportunity to engage in energy sharing and to reduce energy cost without any investment cost. Meanwhile, \cite{Wayes_AE_Mar_2020} proposes a P2P energy sharing technique that enables prosumers to opportunistically use their battery energy for trading when the utility of trading is at a maximum.

\paragraph{Grid benefit}The grid can also exploit P2P energy sharing within the energy network to reduce its peak demand and balancing supply and demand within a community without compromising network security and loss performance \cite{Archie_P2P_Sept_2019,Guerrero_PESGM_2019}. In \cite{Tushar_TSG_Mar_2020}, a cooperative game theory-based hierarchical energy sharing mechanism is proposed in which the grid can interact with a community of prosumers that perform P2P sharing to reduce their peak demand. Similar exploitation of P2P sharing for peak load shaving is also discussed in \cite{Wang_AE_Oct_2019}. For balancing demand and supply, prosumers can be incentivized to employ their distributed resources such as PV and battery storage~\cite{Kirchhoff_AE_June_2019}, coordinate their energy usage, and then buy and sell orders accordingly to balance the demand and supply within a community~\cite{Tushar_TSG_May_2016}. For this purpose, the role of residential storage to provide flexibility is critical~\cite{Luth_AE_Nov_2018}. Examples of P2P mechanisms to balance demand and supply within communities can be found in \cite{Paudel_TIE_Aug_2019} and \cite{Li_AE_Aug_2019}. 
\subsubsection{Community (shared) storage}\label{sec:SharedStorage}Community storage, which is both a technical and social innovation, is expected to contribute positively to building communities primarily using renewable energy resources while accommodating the needs and expectations of the prosumers within local communities~\cite{Koirala_AE_Dec_2018}, while reducing the operating and capital costs of low and medium networks \cite{Ma_AUPEC_2017,Ma_PowerTech_2019}. After a comparison of batteries within 4500 households in 200 communities, it is reported in \cite{Barbour_AE_Feb_2018} that community batteries are more effective in promoting PV integration within communities and promise more benefits compared to household storage systems~\cite{Scheller_AE_July_2020}. A techno-economic study to improve the feasibility of adopting community storage in the UK is reported in \cite{Dong_RSER_Oct_2020}. 

Consequently, the use of community storage in P2P sharing to establish both individual and communal benefits has received much interest in recent times. For example, in \cite{Tarek_ENB_Mar_2017}, a community-based P2P sharing mechanism is proposed in which a centralized entity such as a community manager coordinates the sharing of storage among various prosumers based on their reputations in the reallocation of available energy in the shared storage. Energy management and optimal storage sizing technique for a shared community using a multi-stage stochastic approach is studied in \cite{Hafiz_AE_Feb_2019}. A trilateral planning model for community storage using a bi-level stochastic programming approach is designed in \cite{Kasmaei_TPWRS_Jan_2020} which makes a joint optimal photovoltaic (PV) and energy storage system plan with a view to maximize prosumers benefit. In \cite{Zhong_TSG_EAccess_2020}, a scheme for multi-resource allocation of shared storage is proposed using a distributed combinatorial auction approach with the purpose to reduce the electricity bills of the consumers. Finally, a detailed discussion on equilibrium prices for shared storage in a spot market, technology platforms necessary for the physical exchange of power, and market platforms necessary to trade electricity storage can be found in \cite{Kalathil_TSG_Jan_2019}.

\subsubsection{Mobile storage (electric vehicle)}\label{sec:MobileStorage}Electrification of road transportation is an essential part of alleviating the impact of greenhouse gas on climate change. With much innovations in the design and technicalities of electric vehicles (EV)~\cite{Siang_RSER_Apr_2013,Mwasilu_RSER_Apr_2014}, also known as mobile storage~\cite{Rahmani-Andebili_IET_Apr_2019}, large penetration of EVs on the road have demonstrated the potential to support the grid in terms of providing demand flexibility~\cite{Zhou_ECM_Nov_2019} and frequency regulation~\cite{Peng_RSER_Feb_2017} through their charging and discharging capacities. Additionally, due to the capability of storage mobility, EVs are considered as promising elements for P2P energy sharing~\cite{Alvaro-Hermana_TSM_Fall_2016} where power can be transferred from the battery of one EV to the battery of another EV~\cite{Zhang_TITS_Jan_2019}. The possibility of EVs to be connected together for sharing energy is further enhanced by the recent advancement of artificial intelligence in the EV domain~\cite{Dai_WirelessCommunications_June_2019}.

As a result, several studies have reported different mechanisms to execute P2P energy sharing among EVs. For instance, a distributed EV power trading model based on blockchain and smart contract is proposed in \cite{Liu_Access_vol7_2019} to realize the information equivalence and transparent openness of power trading. A similar technique of smart contracts based proof-of-benefit consensus protocol is also used in \cite{Liu_WirelessCommunication_Feb_2019} to design a P2P energy sharing mechanism for EVs to balance local electricity demand within the community. A multi-objective techno-economic environmental optimization is proposed in \cite{Ridoy_AE_Jan_2020} to enable EVs for energy services such as P2P trading where the EVs much need to cooperate together to achieve the overall social benefit. In \cite{Kang_TII_Dec_2017}, a consortium blockchain-based energy trading mechanism is implemented to enable plug-in EVs to share energy in a localized P2P market. Finally, an operational framework is developed in \cite{Aznavi_TIA_EA_2020} for peer-to-peer (P2P) energy trading between an EV charging station and a business entity equipped with a solar generation unit that outputs significant benefits for both parties compared to having sole agreements with the utility.

\begin{table*}[t]
\centering
\caption{Summary of advancement of P2P energy sharing schemes in different domains of the energy system. Each domain is further divided into sub-domains and an overview of existing studies in those sub-domains are outlined.}
\footnotesize
\begin{tabular}{|m{2cm}|m{2.5cm}|m{9cm}|m{3cm}|}
\hline
\textbf{Domain}&\textbf{Sub-domain} & \shortstack{\textbf{Overview of the sub-domain}} & \textbf{References}\\
\hline
\multirow{3}{*}{\thead{Building}}& Building management & To develop an efficient building management system, which can be utilized to control HVAC, lights, and flexible loads to create energy surplus or reduce energy deficiency in order to contribute to P2P energy sharing. & \cite{Manic_IEM_Mar_2016,RuiJing_AE_Mar_2020,Alam_AE_Mar_2019,Alam_EE_Dec_2017,Ippolito_ENB_Feb_2014,Lazos_RSER_Nov_2014,WenTai_TETC_Sept_2019,Marco_AE_June_2020,PDu_TSG_June_2011,Raja_TETCI_2020,WenTai_TETC_Sept_2019,Ngarambe_ENB_Mar_2020,Valladares_BNE_May_2019,Nagy_ENB_May_2015,Zou_ENB_Jan_2018,Labeodan_ENB_Sept_2016,Liu_ENB_Sept_2016,Li_CSEE_Dec_2018,Georges_AE_Feb_2017,NaveedHassan_TSG_Nov_2015,Habib_TEES_Nov_2016,Wang_TSG_July_2020,Andersen_AE_Dec_2017,Hussain_EJ_June_2018,Stelmach_EP_Sept_2020,Song_Energies_Dec_2019,Zhang_TSG_Nov_2019}\\
\cline{2-4}
&Co-generated electricity and heat management& To develop P2P sharing scheme for households with FC-CHP systems in connected communities to reduce their cost and improving system performance in terms of balancing local supply and demand. & \cite{Wakui_Energy_Feb_2010,Aki_Hydrogen_Nov_2016,Tran_ECM_Oct_2018}\\\cline{2-4}
&Human-centric management& To humanizing the design of HVAC, lighting, and flexible load management based on human behavior and convenience to encourage building owners to participate in P2P sharing. &\cite{Wayes_AE_Mar_2020,Jazizadeh_AE_June_2018,Jung_BE_July_2019,Adhikari_TSG_May_2020,Jiang_EPSR_Sept_2020,Thomas_Nature_2018,Tushar_AE_June_2019,Tushar_Energy_Apr_2020,Tushar_Access_Oct_2018,Paudel_TIE_Aug_2019,Wilkinson_ERSC_Aug_2020} \\\cline{1-4}
\multirow{3}{*}{\thead{Storage}}& Residential storage & To investigate the use of P2P sharing of residential storage to benefit prosumers and the grid. & \cite{Nguyen_AE_Oct_2018,Long_AE_Sep_2018,Jan_EB_Feb_2019,Nizami_AE_Mar_2020,Shantonu_AE_Nov_2018,Rodrigues_AE_Mar_2020,Wayes_AE_Mar_2020,Tushar_TSG_Mar_2020,Wang_AE_Oct_2019,Kirchhoff_AE_June_2019,Tushar_TSG_May_2016,Luth_AE_Nov_2018,Paudel_TIE_Aug_2019,Li_AE_Aug_2019}\\\cline{2-4}
&Community storage&  To understand the use of community storage in P2P sharing to establish both individual and communal benefits in the smart gird. & \cite{Koirala_AE_Dec_2018,Barbour_AE_Feb_2018,Scheller_AE_July_2020,Dong_RSER_Oct_2020,Tarek_ENB_Mar_2017,Hafiz_AE_Feb_2019,Kasmaei_TPWRS_Jan_2020,Zhong_TSG_EAccess_2020,Kalathil_TSG_Jan_2019}\\\cline{2-4}
&Mobile storage& To demonstrate the application of mobile storage, that is EV, to support the grid in terms of providing demand flexibility and frequency regulation and enable prosumers to reduce energy and transport-related costs. & \cite{Siang_RSER_Apr_2013,Mwasilu_RSER_Apr_2014,Rahmani-Andebili_IET_Apr_2019,Zhou_ECM_Nov_2019,Peng_RSER_Feb_2017,Alvaro-Hermana_TSM_Fall_2016,Zhang_TITS_Jan_2019,Dai_WirelessCommunications_June_2019,Liu_Access_vol7_2019,Liu_WirelessCommunication_Feb_2019,Ridoy_AE_Jan_2020,Kang_TII_Dec_2017,Aznavi_TIA_EA_2020}\\\cline{1-4}
\multirow{3}{*}{\thead{Renewable}}& Solar & To design P2P energy sharing schemes focusing on the participation of residential households in reducing energy cost, balancing supply and demand, reducing peak load, and managing network loss. & \cite{JAn_AE_Mar_2020,Wayes_AE_Mar_2020,YJiang_AE_Aug_2020,Paudel_TIE_Aug_2019,Anees_AE_Nov_2019,Long_AE_Sep_2018,Wang_AE_Nov_2019,Andoni_RSER_Feb_2019,Yildiz_AE_Dec_2017,Chen_TSG_July_2019,Gonzalez_IEM_Dec_2018,Tushar_TIE_Apr_2015,Tushar_TSG_Mar_2020,Thomas_Nature_2018,Luth_AE_Nov_2018,Jogunola_Energies_Dec_2017,Tushar_TSG_Jan_2020,Azim_AE_Apr_2020,Nikolaidis_TPWRS_Early_2019,Baroche_TPWRS_July_2019,Xu_TIE_Nov_2019,Thomas_TPS_Early_2018,Thomas_TSG_July_2020}\\\cline{2-4}
&Wind& To develop mechanisms that can be used for wind energy to be shared among participants within connected communities. & \cite{Zhang_AE_June_2018,Dahraie_Systems_EA_2020,Baros_TPWRS_Nov_2017,Arsoon_AE_Mar_2020,Jan_EB_Feb_2019,Ruotsalainen_ERSS_Dec_2017}\\\cline{2-4}
&Hydrogen& To explore the opportunities for P2P hydrogen sharing in the energy network that includes fuel cell vehicles, energy storage, combined heat and power system, and renewable energy. & \cite{LeiLi_RSER_Apr_2019,Robledo_AE_Apr_18,Zhu_AE_Aug_2020,Hasan_RE_Aug_2020,Xiao_TPWRS_July_2018}\\\hline
\end{tabular}
\label{table.ref.domains}
\end{table*}

\subsection{Renewable domain}\label{sec:RenewableDomain}In general, solar, wind, and hydrogen are becoming popular to produce renewable energy for usage within the community. The benefits of using renewable energy for P2P sharing such as cost reduction, demand-supply balance, and peak reduction are well established. However, to enjoy these benefits, at least some prosumers within the connected community should be able produce surplus energy to be shared by other participants. Further, sharing renewable energy in P2P decentralized market without any coordination can compromise the operation of the network within its technical limit. Hence, prosumers' transactive meter should have the capacity to decide how much energy it can push to the P2P network at any given time without compromising the network security. Nonetheless, sharing renewable energy is the most popular and well investigated P2P sharing mechanism in the literature. It offers benefit to prosumers to enjoy low cost clean energy without investing in energy storage. Meanwhile, the advancement in inverter technology has further opportunity to benefit prosumers economically by enabling them to provide regulation services to the grid. In this section, we discuss existing P2P sharing techniques for three different renewable sources: solar, wind, and hydrogen.

\subsubsection{Solar}\label{sec:Photovoltaic}As discussed before, most existing literature on P2P energy sharing focuses on the participation of residential households, which are usually equipped with rooftop solar panels. As a result, in the renewable domain, most P2P sharing mechanisms are developed considering solar as the main source of energy. In particular, the use of solar in P2P trading has been extensively used for reducing energy cost, balancing supply and demand, reducing peak load, and managing network losses.
\paragraph{Cost reduction}Reducing energy cost is possibly the most important characteristic of P2P sharing that encourages prosumers to install rooftop solar and participate in the local energy market. However, how much cost prosumers can save relies on the energy sharing price and strategy of each participating prosumer~\cite{JAn_AE_Mar_2020}. For example, in \cite{Wayes_AE_Mar_2020} the authors proposed an opportunistic energy sharing mechanism using solar panel and batteries that helps prosumers to reduce their energy costs. A multi-leaders and multi-followers based P2P sharing model is proposed in \cite{YJiang_AE_Aug_2020} that reduces the cost of buyers by 4.36\% while improves the benefit of sellers by 12.61\% compared to the feed-in-tariff scheme. It is reported in \cite{Paudel_TIE_Aug_2019} and \cite{Anees_AE_Nov_2019} that extensive engagement of prosumers in P2P sharing is the key to reduce energy cost, which can further be improved by additional inclusion of energy storage devices~\cite{Long_AE_Sep_2018,Wang_AE_Nov_2019}.
\paragraph{Supply-demand balance}Within a connected community, it is critical that some prosumers have energy generation capacity to enable prosumers to participate in energy sharing with one another. Now, to reap the maximum economic benefit, it is important that the local generation balances the local demand, which is generally monitored through distributed ledger techniques for P2P sharing~\cite{Andoni_RSER_Feb_2019}. For example, in P2P sharing, prosumers can monitor their own energy generation and demand through smart meter~\cite{Yildiz_AE_Dec_2017}, access the energy offered by other prosumers for sharing within the market through distributed ledger~\cite{Chen_TSG_July_2019}, and then subsequently create the buy and sell order to share with one another within the community~\cite{Gonzalez_IEM_Dec_2018} to balance the supply and demand. However, if there is a deficiency of energy, it can be supplied by the grid~\cite{Tushar_TIE_Apr_2015}, community storage~\cite{Barbour_AE_Feb_2018}, or diesel generators~\cite{Zhang_TIA_July_2016} with relatively higher cost.
\paragraph{Coordinated peak demand reduction}An important service that can be provided by prosumers to the grid by participating in P2P sharing is peak demand reduction. Several studies have reported techniques that have proven effective to reduce peak demand. For example, in \cite{Tushar_TSG_Mar_2020}, the authors propose a cooperative game theory-based P2P sharing scheme that helps a centralized power system to reduce the total electricity demand of its customers at peak hours. A federated power plant using P2P energy sharing platforms is introduced in \cite{Thomas_Nature_2018}, which incentivizes prosumers to shift their loads aware of predictable peak demand periods. In \cite{Luth_AE_Nov_2018}, the authors show how adjusting the structure of demand tariffs can reduce their contribution to peak demand. Further applications of P2P energy sharing in reducing peak demand can be found in \cite{Jogunola_Energies_Dec_2017}. 
\paragraph{Network loss management}P2P energy sharing can incur additional power loss within the system due to the transacted electrons by different prosumers to the network~\cite{Tushar_TSG_Jan_2020}. For example, the impact of P2P sharing on the change in network power loss is discussed in \cite{Azim_AE_Apr_2020}. The cost of this network loss needs to be distributed. As such, in \cite{Nikolaidis_TPWRS_Early_2019}, the authors propose a graph-based scheme for allocating cost among the participating prosumers. Fees are used by the system operator in \cite{Baroche_TPWRS_July_2019} to allocate market-related grid costs to the participants. Further, an optimal power routing strategy is used in \cite{Xu_TIE_Nov_2019} to optimize the power dispatch by different prosumers with the objective of minimizing the power loss ratio between the buyers and sellers within the network, while\cite{Thomas_TPS_Early_2018} introduces energy classes to treat energy as a heterogeneous product and coordinate P2P sharing to minimize the cost of network losses. As an alternative, \cite{Guerrero_PESGM_2019} investigate how a P2P trading mechanism can be designed to bias the matching of electrically close peers, in order to reduce losses.
\subsubsection{Wind}\label{sec:Wind}Compared to solar, the number of studies covering P2P sharing of wind energy is relatively small. This is mainly due to the fact that residential houses usually do not install wind turbines to produce onsite energy. Wind turbines are generally used as small or medium scale wind farms within microgrids~\cite{Zhang_AE_June_2018}. However, some mechanisms have been reported that can be used for wind energy to be shared among participants within connected communities. For example, in \cite{Dahraie_Systems_EA_2020}, a stochastic decision-making framework is designing in which a wind power generator can provide some required reserve capacity from demand response aggregators in a P2P structure. The proposed framework is formulated as a bilevel stochastic model incorporating the capability to assess risk associated with wind power generator's decision and to determine the effect of scheduling reserves on the profit variability.

In \cite{Baros_TPWRS_Nov_2017}, a distributed control methodology is proposed to control the power outputs of deloaded wind double-fed induction generators under dynamical conditions and through peer-to-peer information exchange. P2P energy trading for enhancing the resilience of networked microgrids with wind turbines is studied in \cite{Arsoon_AE_Mar_2020}. After defining a community of houses containing a mix of prosumers (with a PV and/or wind power installation) and consumers, a two-stage stochastic model of P2P sharing model is proposed in \cite{Jan_EB_Feb_2019}. The authors have verified the proposed approach and assessed its economic potentials. Lastly, a critical P2P vision of renewable energy including wind energy to describe the relationship between energy transitions and social change, and to offer one plausible socio-cultural vision of the era of renewable energy is discussed in \cite{Ruotsalainen_ERSS_Dec_2017}.
\subsubsection{Hydrogen}\label{sec:Hydrogen}With the establishment of the Paris agreement on November 4, 2016, it is expected that the utilization of hydrogen will not only enhance the sustainability and reliability of the energy system but also improve the system's flexibility~\cite{LeiLi_RSER_Apr_2019}. While today's energy sector is heavily dependent on fossil fuel, hydrogen can play a pivotal role in the future to establish a low carbon economy by linking various layers of energy transmission and distribution infrastructure~\cite{LeiLi_RSER_Apr_2019}. For example, to improve the absorption of renewable energy, hydrogen is already used for small scale residential storages (fuel cells) and transportations. In \cite{Robledo_AE_Apr_18}, the results from a demonstration project on a zero-energy residential building are discussed, which consist of both solar panels and hydrogen fuel cell EV for combined transport and power generation. According to \cite{Robledo_AE_Apr_18}, the project has demonstrated a reduction of energy import from the grid by 71\% over one year, which is an excellent outcome towards achieving a net-zero carbon energy system.

As hydrogen has begun to penetrate to the energy system, researchers have also begun to explore the opportunities for P2P hydrogen sharing in the energy network. In \cite{Zhu_AE_Aug_2020}, a multi-agent management framework including fuel cell vehicles, energy storage, combined heat and power system, and renewable energy is proposed for energy sharing among multiple microgrids. The main purpose is to schedule the arrangements of fuel cell vehicles to improve the local absorption of renewable energy and enhance the economic benefits of microgrids. A P2P home energy sharing technique is studied in \cite{Hasan_RE_Aug_2020} to find the optimal size and operational schedule for hydrogen storage  and solar systems incorporated within homes. In addition, \cite{Xiao_TPWRS_July_2018} develops a local energy market framework, in which electricity and hydrogen are shared. Participants in the market consist of renewable distributed generators, loads, hydrogen vehicles, and a hydrogen storage system operated by an agent and an iterative market-clearing method is designed where participants submit offers/bids with consideration of their own preferences and profiles according to the utility functions. It is shown that the proposed clearing process avoids complex calculations and preserves players' privacy.

A summary of the advancement of P2P energy sharing in different domains of the energy network is given in Table~\ref{table.ref.domains}.
\section{Pilot projects around the world}\label{sec:PilotProjects}To demonstrate the effectiveness of P2P energy sharing in the smart grid, a large number of pilot projects are being trialed in different parts of the world. In particular, countries in North America, Europe, Australia, and Asia are heavily engaged in studies under various testbed settings. To that end, what follows is an overview of different projects on P2P energy sharing in four continents of the world. For each continent, we provide some detail of one particular project and followed by a summary of other projects on P2P sharing.

\subsubsection{North America}In North America, almost all P2P projects are based in the USA. For example, a P2P project can be found in Brooklyn in the Brooklyn microgrid testbed, in which prosumers within a community can trade their on-site produced energy with the neighbors by using the typical power network~\cite{Esther_AE_Jan_2018}. The Brooklyn Microgrid project consists of a microgrid energy market in Brooklyn, New York, where participants are located across three distribution grid networks. Severe weather events (e.g. hurricanes, heat waves, etc.) increase, which raises operation issues of the (already) outdated electrical grids in Brooklyn. In this project, the consumption and generation data is transferred from the prosumers Transactive Grid smart meters to their blockchain accounts to create buy and sell orders. Orders are sent to the market mechanism which is sustained by a smart contract. When buy and sell orders match, payment is carried out and a new block with all current market information is added to the blockchain. This microgrid is helping the grid to reduce the impact of grid issues through complete decoupling and control the energy supply within the community in the event of natural calamity. Further, it is also helping the grid to accommodate the growing number of electric vehicles on the US road~\cite{Esther_AE_Jan_2018}.

Among other projects, a cloud-based software platform is used in the project TeMiX \cite{TeMiX2020} for decentralized network management purposes, which also enabled automated energy transactions between peers within the community. Similar solar implementation and software platform for trading surplus solar generation between houses was trialed in Boston \cite{Yeloha2020}. Another new project on P2P sharing is going to begin in PowerNet's headquarter in Florida, in which Power Ledger's xGrid platform will be used to trade  solar power with neighbors connected in its office park~\cite{PowerNet2020}. 
\begin{table*}[t]
\centering
\caption{Summary of different projects on P2P energy sharing in North America and Europe.}
\footnotesize
\begin{tabular}{|m{2cm}|m{2.5cm}|m{9cm}|m{3cm}|}
\hline
\textbf{Continent}&\textbf{Country} & \shortstack{\textbf{Overview of the project}} & \textbf{References}\\
\hline
\multirow{4}{*}{\thead{North America}}& \multirow{4}{*}{\thead{USA}} & A P2P energy sharing demonstration project in Brooklyn using the blockchain where prosumers share their energy using typical distribution network & \cite{Esther_AE_Jan_2018}\\
\cline{3-4}
&& A cloud-based software platform TeMiX for energy trading that provide automated energy transaction and decentralized network management services & \cite{TeMiX2020}\\\cline{3-4}
&& Kealoha project has implemented P2P markets using solar generation where a software platform enable exchange of excess solar generation between houses &\cite{Yeloha2020} \\\cline{3-4}
&&Power Ledger's xGrid platform will be used in American PowerNet's headquarters to trade solar power with neighbors connected in its office park &\cite{PowerNet2020}\\\cline{1-4}
\multirow{8}{*}{\thead{Europe}}& \multirow{3}{*}{\thead{Germany}} & Share\&Charge is a decentralized blockchain based market for EV charging, transactions, and data sharing & \cite{ShareAndCharge2020}\\\cline{3-4}
&& Peer energy cloud is a cloud based local energy platform for local energy trading and smart homes &\cite{PeerEnergyCloud2020} \\\cline{3-4}
&&Sonnen community considers storage and storage systems to provide a platform for virtual energy pool &\cite{sonnen2020}\\\cline{2-4}
&\multirow{2}{*}{\thead{Netherlands}} & Powerpeers is a blockchain based market that has enabled residential buildings to share energy with one another &\cite{Powerpeer2020}\\\cline{3-4}
&& Vandebron helps electricity consumers to select local sustainability produces of their choices for energy purchase&\cite{vendebron2020}\\\cline{2-4}
&\textbf{Norway} & EmPower is a local energy trading platform in which prosumers can share their energy with one another in a local energy market&\cite{Empower2017}\\\cline{2-4}
&\textbf{UK} &Piclo is a software platform for selling and buying smart grid flexibility services and trading energy among peers &\cite{Piclo2020}\\\cline{2-4}
&\textbf{Finland} &SmartTest is a smart energy system with the consideration of information and communication technology and P2P sharing of resources  &\cite{SmartTest2020}\\\hline
\end{tabular}
\label{table.ref.USA.Europe}
\end{table*}
\subsubsection{Europe}Based on the number of projects on P2P sharing, undoubtedly Europe is leading the world with a number of demonstration trials in Germany, Netherlands, Norway, Finland, and the UK. For instance, three projects on P2P sharing are available in Germany. \emph{Share\&Charge} is a blockchain energy market for EV charging transactions, and data sharing~\cite{ShareAndCharge2020}. \emph{Peer energy cloud}~\cite{PeerEnergyCloud2020}  is a cloud-based platform that enable local energy sharing by considering Information and communication technology and P2P approaches. Sonnen Community in Germany considers solar and storage systems to create a virtual energy pool~\cite{sonnen2020}. Meanwhile, in the Netherlands, two projects on P2P sharing are running at this moment including 1) a project - known as \emph{Powerpeers} - for residential buildings to share their energy with one another using a blockchain-based energy market~\cite{Powerpeer2020} and 2) \emph{Vandebron} - a platform for electricity consumers to select desirable local sustainability producers~\cite{vendebron2020}. In Norway, \emph{EMPower} provides a trading platform for local energy exchange between prosumers in a local market~\cite{Empower2017}, whereas \emph{Piclo} is a UK based software platform for selling and buying of smart grid flexibility services and P2P energy trading~\cite{Piclo2020}. 

Lastly, \emph{P2P-SmartTest} in Finland demonstrates a smart grid based transactive energy concept to perform P2P energy sharing~\cite{SmartTest2020}. The objective of the P2P-SmartTest project is to investigate and demonstrate a smarter electricity distribution system integrated with advanced ICT, regional markets, and innovative business models. The developed P2P approaches ensure the integration of demand-side flexibility and the optimum operation of DER and other resources within the network. Meanwhile, they are also capable of maintaining a second-to-second power balance and the quality and security of the supply. A part of the project also develops and demonstrates the capability of the distributed wireless ICT solutions in offloading the required traffic of different applications of energy trading, network optimization, and real-time network control. For proper operation of the distributed network, the project integrates the necessary network operation functions for resilient distribution system operation. Table~\ref{table.ref.USA.Europe} summarized the projects on P2P sharing in North America and Europe.
\subsubsection{Australia}With extensive government subsidy from both Federal and State governments, for example, see \cite{ARENA_P2P_2020}, P2P energy sharing has also gotten significant momentum in Australia. With the establishment of Power Ledger, in particular, P2P energy sharing pilots are being demonstrated in several areas in Australia. For example, in RENew Nexus project, Power Ledger has conducted trials with different households, where the households trade excess energy generated from rooftop solar panels with their with neighbors using the existing electricity network and retailers~\cite{RENeW2020}. It is an on-going project in Western Australia with the purpose to understand the potential of localized energy markets and how technology platforms can facilitate more efficient outcomes to the energy system. The project ran in three parts. In part 1 - Freo 48 - a solar P2P trial was run in two phases. Phase 1 ran for seven months in 2018-19 with 18 participants whereas Phase 2 initially had 30 participants (later reduced to 29) and ran from October 2019 to January 2020. In part 2 - Loco 1 - the modeling of a residential Virtual Power Plant (VPP) was completed to better understand the financial benefits that prosumers could realize from having a battery installed and participating in a VPP. Finally, part 3 - Loco 2 - a microgrid with a 670kWh shared battery system which will facilitate 36 households is currently under construction. This will trade excess energy with each other via the battery. The purpose of these trials was (1) to demonstrate proof of concept test for P2P electricity trading; (2) to understand the value of P2P for customers and project partners; and (3) to trial the technological interoperability between the Power Ledger platform, Synergy (energy retailer), Western Power (network operator) and supporting technologies, and test the Power Ledger platform capability as a client P2P solution. Other examples of P2P projects in Australia are explained in Table~\ref{table.ref.Australia.Asia}.
\begin{table*}[t]
\centering
\caption{Summary of different projects on P2P energy sharing in Australia and Asia.}
\footnotesize
\begin{tabular}{|m{2cm}|m{2.5cm}|m{9cm}|m{3cm}|}
\hline
\textbf{Continent}&\textbf{Country} & \shortstack{\textbf{Overview of the project}} & \textbf{References}\\
\hline
\multirow{9}{*}{\thead{Australia}}& \multirow{10}{*}{\thead{Australia}} & Power Ledger partnered with Nicheliving to deploy its energy trading platform to deliver 100\% renewable energy at 62 apartments & \cite{Nicheliving2020}\\
\cline{3-4}
&& An Australian retail investment company uses Power Ledger's content to trial trading solar energy within its shopping centers &\cite{Retail2020} \\\cline{3-4}
&&Energy retailer Powerclub will use Power Ledger's VPP enabled platform to enable Powerclub households with batteries to sell their stored solar energy during peak demand period and price spikes &\cite{Powerclub2020}\\\cline{3-4}
&& RENeW Nexus involved two trials whereby 48 households used Power Ledger's platform to trade excess energy generated from rooftop solar panels with their neighbours via the existing electricity network and retailer &\cite{RENeW2020}\\\cline{3-4}
&& 36 homes with rooftop solar in East Village, WA will use Power Ledger's platform to trade solar energy with each other via the battery&\cite{EastVillage2020}\\\cline{3-4}
&& The Gen Y Demonstration Housing Project in WA uses Power Ledger's blockchain platform to sell excess solar energy to neighbors at peak demand, rather than back to the grid&\cite{GenY2020}\\\cline{3-4}
&&In Wongan-Ballidu, Australia nine commercial sites will use Power Ledger's P2P energy trading platform to monetize their excess solar energy &\cite{Wongan2020}\\\cline{3-4}
&&The first P2P commercial project in the National Energy Market in the Australian Capital Territory, which  will enable the customer to save on their energy costs and decrease consumption from fossil fuel sources  &\cite{EPCSolar2020}\\\cline{3-4}
&&In DeHavilland Apartments \& Element47, Australia, Power Ledger will have its peer-to-peer (P2P) energy trading technology will be used to enable households to trade solar energy between one another  &\cite{DeHavilland2020}\\\cline{1-4}
\multirow{11}{*}{\thead{Asia}}& \multirow{3}{*}{\thead{Japan}} & Kansai Electric Power Co (Kepco) is leading a project that will enable solar power suppliers to sell extra electricity to consumers via a blockchain-enabled system & \cite{Japan_1_Pilot}\\
\cline{3-4}
&& KEPCO used Power Ledger's platform to create, track, trade and provide a marketplace for the settlement of renewable energy credits or non-fossil certificates, generated by rooftop solar systems &\cite{Japan_2_Pilot} \\\cline{3-4}
&& Blockchain-enabled peer-to-peer technology demonstration with households trading excess solar energy between each other in the Kanto region &\cite{Japan_3_Pilot} \\\cline{2-4}
&\multirow{2}{*}{\thead{India}} & A peer-to-peer pilot energy trading project in India's Lucknow will demonstrate the feasibility of Power Ledger's platform to trade energy from rooftops with solar power to neighboring households/buildings &\cite{India_1_Pilot} \\\cline{3-4}
&& A large-scale desktop P2P energy trading trial across existing 300 kW solar infrastructure servicing a group of gated communities in Delhi's Dwarka region &\cite{India_2_Pilot} \\\cline{2-4}
&\multirow{2}{*}{\thead{Thailand}} & Power Ledger, in partnership with TDED, will create a blockchain-based digital energy business developing peer-to-peer energy trading solutions in Thailand &\cite{Thailand_1_Pilot} \\\cline{3-4}
&& Thai renewable energy business BCPG and Thai utility Metropolitan Electricity Authority are using Power Ledger's software for tracking and settling the energy generated from renewable sources and transactions between the participants &\cite{Thailand_2_Pilot} \\\cline{2-4}
&\multirow{2}{*}{\thead{South Korea}} & Electron, a blockchain startup in UK, will test their own energy flexibility trading platform in South Korea to prove the benefits of flexibility trading in the South Korean market as it starts to decarbonize and decentralize.&\cite{SKorea_1_Pilot}\\\cline{3-4}
&& KEPCO will trial a blockchain-based peer-to-peer energy trading system in two apartments in Seoul and nine buildings within KEPCOs facilities to lower energy bills by enabling businesses and households to buy surplus electricity of their peers&\cite{SKorea_2_Pilot}\\\cline{2-4}
&\textbf{Singapore} &Electrify's Marketplace 2.0 uses smart contracts where consumers can buy electricity from retailers or even from their own peers with the purpose to eliminate or reduce many of the fees and transaction costs with the automatic nature of smart contracts &\cite{Singapore_1_Pilot}\\\cline{2-4}
&\textbf{Malaysia} & A pilot trial aims to demonstrate the feasibility of solar energy trading in the Malaysian energy market  by enabling consumers to choose whether they wish to purchase clean, renewable energy or power from fossil fuels &\cite{Malaysia_2_Pilot}\\\hline
\end{tabular}
\label{table.ref.Australia.Asia}
\end{table*}
\subsubsection{Asia}Asia is also not behind when it comes to demonstrate the effectiveness of P2P energy sharing for energy producers and consumers. For example, in Japan, the Australia-based Power Ledger made a partnership with Japanese solar provider Sharing Energy and electricity retailer eRex to trial its P2P trading platform in Kanto, Japan~\cite{Japan_2_Pilot}. The trial aimed to show how a group of households can trade excess solar energy between each other and follows a trial in Osaka with Japan's utility KEPCO which achieved consumer acceptance. In the trial, Power Ledger's platform was integrated with existing smart meter systems in homes to enable participants to set prices and track energy trading in real-time. The trial was scheduled to run until December 2019.

Although Japan, Thailand, South Korea, and India are leading the efforts with more than one demonstration project, Singapore and Malaysia are also in the race to contribute towards P2P demonstration through  different pilots on distributed energy sharing. A summary of different projects of P2P trading that are currently being trialed in Asia is given in Table~\ref{table.ref.Australia.Asia}.
Note that, except P2P sharing, other relevant energy trading techniques such as transactive control~\cite{Rahimi_EM_June_2018} and negawatt trading~\cite{Tushar_NE_2020} are also trialed in many parts of the world. For details of these demonstration projects, readers are referred to \cite{Abrishambaf_ESR_Nov_2019}, \cite{Andoni_RSER_Feb_2019}, and \cite{Okawa_Negawatt_Nov_2017}.
\section{Summary \& Challenges}\label{sec:challenges}Based on the discussion in this paper, clearly, the benefit of P2P sharing to both prosumers and the grid is well demonstrated. Prosumers can reduce the cost of electricity while enjoying clean energy by participating in P2P sharing.  By encouraging community members to interact with one another to share their energy and introducing provisions for energy donation, P2P sharing improves social values within the members and help them to establish an environment-friendly energy neighborhood. Previously, for example, in Feed-in-Tariff only prosumers with renewable sources could enjoy the economic and environmental benefits of energy trading. However, with P2P, prosumers with any renewable energy status can have that luxury. Now, in some P2P markets, prosumers can make their decision on not only the energy sharing parameters but also whether or not to participate in energy sharing without any influence from a third party. Thus, P2P empowers the prosumers and give them the true independence of the energy they produce and manage.

Meanwhile, the grid can also benefit significantly from P2P sharing. For instance, by reducing the demand, P2P sharing can help the grid to reduce its investment for infrastructure upgrades to cope with the increasing energy demand. This also opens the opportunity for the grid to accommodate new emerging demand for energy from the electric vehicle with its current infrastructure with a minimal upgrade. Now, building within a connected community can provide demand flexibility to the grid through which the grid is able to reduce its peak load demand and improve the security of the power system by addressing the voltage and frequency disturbances through the ancillary service market.

However, despite demonstrated benefit and ample opportunities, the establishment of P2P sharing in today's energy market relies on the decision of the regulatory board. That is, the regulatory board determines how P2P energy markets fit into the current energy policy. Thus, legislative rules establish which market design is allowed, how taxes and fees are designed, and in which way the market can be integrated into today's energy market and energy supply system. Thus, the government of country/region can easily support P2P markets to enable the efficient utilization of local renewable energy resources and decrease environmental degeneration by regulatory changes, e.g. the introduction of subsidies. The regulatory board can also discourage the implementation of P2P markets if they determine that the subsequent result could have negative impacts on the current energy system.

Meanwhile, to present P2P sharing to the regulatory board as a viable energy management option and convince them to allow it to practice in the current energy market, a number of potential challenges that are yet to be addressed, These challenges can be summarized as follows.

\subsection{Co-existence of different stakeholders}Of course, the research in P2P energy sharing has been extensive for settings where prosumers are engaged in sharing energy among themselves with very little (or, not at all) with little (or, no) interaction with the grid or an aggregator. However, while this assumption is valid for the establishment of sharing mechanisms in a small-scale setting, to successfully achieve objectives of higher monetary benefit, supply-demand balance, incentivizing prosumers, and ensuring secure transactions, large-scale development of energy sharing requires to consider other existing participating stakeholders of the network. Examples of such stakeholders may include generators, retailers, and distribution network service providers (DNSP).

Now, the involvement of stakeholders with conflicting interests in using prosumers' energy makes the decision-making process of energy sharing a complex task. For example, a conflicting scenario may arise between a retailer and a DNSP serving the same set of prosumers when the retailer requests prosumers to discharge their batteries to meet the excess demand of its additional customers, whereas, at the same time, the DNSP may send a signal to prosumers not to push any electron to avoid a potential network voltage violation expected by the DNSP. Hence, there is need to develop techniques that will prepare P2P energy sharing mechanisms for implementation in systems where stakeholders like retailers, generators, and DNSPs co-exist with other network elements without affecting each other's interests and roles in the network.
\subsection{Network constraints, losses, and management}As the number of prosumers participating in P2P energy sharing will increase, the risk of voltage rise at various nodes of the power system network will increase as well. Of course, one potential way to mitigate this risk could be to regulate how much each prosumer can export to the network at any given time slot. However, imposing such rigid restrictions on prosumers' choices may detrimentally affect the potential revenue that a prosumer would expect from participation in the trading and therefore could cause prosumers to lose interest in future participation. On the other hand, unchecked injection of power from all prosumers will compromise the security of the network. 

Another important issue is that P2P transactions may raise power losses across the network. Therefore, the P2P price, which has been assumed to be significantly more lucrative than other existing trading schemes in most existing studies will need to consider this loss factor. This will increase the price of energy buyers. Furthermore, a large number of renewable energy plants are being connected to the network, which will allow the retailers to sell energy at a much cheaper rate than before, which may also affect the P2P price as well. Hence, how to allocate the loss factor within P2P price while simultaneously being competitive need to address.   

Furthermore, it is important to note that the flow of electricity cannot be controlled. Therefore, it is not likely that for a very large network the intended receiver will receive the actual power that has been sent to it by the seller over the distribution network. As a consequence, the power loss due to P2P sharing trading would be different compared to the case when the buyer actually receives the power sent by the seller (e.g. if they are located side-by-side). Hence, how to calculate the actual loss in this kind of scenario and then decide the P2P energy sharing price needs further investigation.
\subsection{Post-settlement uncertainty}Although P2P sharing can offer significant benefits to its customer in terms of different energy services and lower energy cost, the outcome of the sharing could jeopardize the overall market structure and trust between the buyers and sellers if the interaction and negotiation between prosumers are poorly designed due to communication delay, insufficient forecasting, less visibility and understanding of network condition, and lack of customer information. For instance, assume a scenario where a prosumer commits a certain amount of energy for a given time slot and gets paid for the committed energy from the seller. However, due to a lack of accurate forecasting of its own demand and generation, it is possible that the actual energy that is transferred to the buyer is less than what was committed. Such a market outcome will surely be suboptimal and not suitable for long-term sustainability.

Hence, before deploying in the energy market, the computation and communication complexity issues must be resolved for the robust operation of the system.  Accurate forecasting techniques need to be employed and network conditions for each P2P transaction periods need to be identified through sophisticated algorithm. Importantly, prosumers need to be educated about the importance of P2P sharing and the discloser of honest information about their demand, generation, and energy commitment to the market for the sustainability of the market.
\subsection{Low cost privacy and security}Finally, one significant importance of P2P energy sharing is the availability of accurate and statistically useful energy transaction and usage data across communities for better prosumer decision-making. However, such data sharing could potentially compromise the privacy of individual participants. Therefore, accessible data also needs to ensure that everyone's private information is safe. For example, a provably-private transformation of prosumers energy data is needed to facilitate data accessibility while simultaneously granting data enough statistical accuracy for interrogation of data.

Furthermore, although P2P energy sharing provides a platform to engage individual prosumers to share their energy with one another and with the grid, without cryptography, this could also increase the vulnerability of the network if one or more prosumers act as adversaries and plan to breach the security of the system, for example, by injecting false information. At present, the security of transactions in P2P energy sharing has been confirmed through the use of distributed ledger techniques such as blockchain. However, it is important to note that providing security via blockchain could be computationally expensive and therefore very costly~\cite{Fairley_Spectrum_Oct_2017}. As such, there is a need for methodologies and techniques that will confirm the security of the network, not by integrating expensive measures, but enabling the trading decision in such a way that the security of the transaction is preserved.
\section{Conclusion}\label{sec:conclusion}This review article has provided an overview of existing peer-to-peer energy sharing literature in order to identify recent progress in this area of research as well as to discover the challenges that are preventing peer-to-peer sharing to become a viable energy management option in today's electricity market. As such, first, we have added background on the connected community, peer-to-peer energy sharing systems, peer-to-peer energy markets, and distributed ledger technology for readers to easily follow the rest of the discussion in the paper. Then, we have discussed recent advancements in peer-to-peer energy sharing research in different domains including building domain, storage domain, and renewable domain. Following the discussion on advancement, a detailed list of existing trial projects on peer-to-peer energy sharing in North America, Europe, Australia, and Asia has been provided. Finally, we have identified a number of challenges that are need to be addressed before deploying peer-to-peer sharing as an energy management technique in today's electricity market.

\section*{Acknowledgement}This work was supported in part by the Queensland State Government under the Advance Queensland Research Fellowship AQRF11016-17RD2, in part by the University of Queensland Solar (UQ Solar; solar-energy.uq.edu.au), and in part by the SUTD-MIT International Design Centre (idc; idc.sutd.edu.sg), in part by the U.S. National Science Foundation under Grants DMS-1736417 and ECCS-1824710, and in part by the Engineering and Physical Sciences Research Council (award references EP/S000887/1 and EP/S031901/1).

\begin{thebibliography}{100}
\providecommand{\url}[1]{#1}
\csname url@samestyle\endcsname
\providecommand{\newblock}{\relax}
\providecommand{\bibinfo}[2]{#2}
\providecommand{\BIBentrySTDinterwordspacing}{\spaceskip=0pt\relax}
\providecommand{\BIBentryALTinterwordstretchfactor}{4}
\providecommand{\BIBentryALTinterwordspacing}{\spaceskip=\fontdimen2\font plus
\BIBentryALTinterwordstretchfactor\fontdimen3\font minus
  \fontdimen4\font\relax}
\providecommand{\BIBforeignlanguage}[2]{{%
\expandafter\ifx\csname l@#1\endcsname\relax
\typeout{** WARNING: IEEEtran.bst: No hyphenation pattern has been}%
\typeout{** loaded for the language `#1'. Using the pattern for}%
\typeout{** the default language instead.}%
\else
\language=\csname l@#1\endcsname
\fi
#2}}
\providecommand{\BIBdecl}{\relax}
\BIBdecl

\bibitem{SolarReport2020}
\BIBentryALTinterwordspacing
A.~E. Council, ``{Solar Report: January 2020},'' Australian Energy Council,
  Feb. 2020, accessed on June 17, 2020. [Online]. Available:
  \url{https://www.energycouncil.com.au/media/18020/australian-energy-council-solar-report_-jan-2020-final.pdf}
\BIBentrySTDinterwordspacing

\bibitem{SolarPV2030}
G.~Parkinson, ``Australia rooftop solar installs total 2.13gw in 2019 after
  huge {December} rush,''
  url={https://reneweconomy.com.au/australia-rooftop-solar-installs-total-2-13gw-in-2019-after-huge-december-rush-34613/},
  Jan. 2020.

\bibitem{Xu_TPWRS_May_2020}
H.~{Xu}, A.~D. {Dom\'{i}nguez-Garc\'{i}a}, V.~V. {Veeravalli}, and P.~W.
  {Sauer}, ``Data-driven voltage regulation in radial power distribution
  systems,'' \emph{IEEE Transactions on Power Systems}, vol.~35, no.~3, pp.
  2133--2143, May 2020.

\bibitem{Scott_TSG_Nov_2019}
P.~{Scott}, D.~{Gordon}, E.~{Franklin}, L.~{Jones}, and S.~{Thi\'{e}baux},
  ``Network-aware coordination of residential distributed energy resources,''
  \emph{IEEE Transactions on Smart Grid}, vol.~10, no.~6, pp. 6528--6537, Nov.
  2019.

\bibitem{Luth_AE_Nov_2018}
A.~L\"{u}th, J.~M. Zepter, P.~C. del Granado, and R.~Egging, ``{Local
  electricity market designs for peer-to-peer trading: The role of battery
  flexibility},'' \emph{Applied Energy}, vol. 229, pp. 1233--1243, Nov. 2018.

\bibitem{Thomas_Nature_2018}
T.~{Morstyn}, N.~{Farrell}, S.~J. {Darby}, and M.~D. {Mcculloch}, ``Using
  peer-to-peer energy-trading platforms to incentivize prosumers to form
  federated power plants,'' \emph{Nature Energy}, vol.~3, no.~2, pp. 94--101,
  2018.

\bibitem{Tushar_TIE_Apr_2015}
W.~{Tushar}, B.~{Chai}, C.~{Yuen}, D.~B. {Smith}, K.~L. {Wood}, Z.~{Yang}, and
  H.~V. {Poor}, ``Three-party energy management with distributed energy
  resources in smart grid,'' \emph{IEEE Transactions on Industrial
  Electronics}, vol.~62, no.~4, pp. 2487--2498, Apr. 2015.

\bibitem{Tushar_AE_June_2019}
W.~Tushar, T.~K. Saha, C.~Yuen, T.~Morstyn, M.~D. McCulloch, H.~V. Poor, and
  K.~L. Wood, ``A motivational game-theoretic approach for peer-to-peer energy
  trading in the smart grid,'' \emph{Applied Energy}, vol. 243, pp. 10--20,
  June 2019.

\bibitem{Imran_PESGM_2020}
M.~I. Azim, W.~Tushar, and T.~Saha, ``{Regulated P2P energy trading: A typical
  Australian distribution network case study},'' in \emph{IEEE PES General
  Meeting (GM)}, Montreal, Canada, Aug. 2020, pp. 1--5.

\bibitem{Arnold_TPWRS_Jan_2018}
D.~B. {Arnold}, M.~D. {Sankur}, M.~{Negrete-Pincetic}, and D.~S. {Callaway},
  ``Model-free optimal coordination of distributed energy resources for
  provisioning transmission-level services,'' \emph{IEEE Transactions on Power
  Systems}, vol.~33, no.~1, pp. 817--828, Jan. 2018.

\bibitem{Tushar_TSG_Jan_2020}
W.~{Tushar}, T.~K. {Saha}, C.~{Yuen}, D.~{Smith}, and H.~V. {Poor},
  ``Peer-to-peer trading in electricity networks: {An} overview,'' \emph{IEEE
  Transactions on Smart Grid}, vol.~11, no.~4, pp. 3185--3200, July 2020.

\bibitem{Archie_P2P_Sept_2019}
J.~{Guerrero}, A.~C. {Chapman}, and G.~{Verbi\v{c}}, ``Decentralized {P2P}
  energy trading under network constraints in a low-voltage network,''
  \emph{IEEE Transactions on Smart Grid}, vol.~10, no.~5, pp. 5163--5173, Sept.
  2019.

\bibitem{Zhang_TSG_July_2020}
K.~{Zhang}, S.~{Troitzsch}, S.~{Hanif}, and T.~{Hamacher}, ``Coordinated market
  design for peer-to-peer energy trade and ancillary services in distribution
  grids,'' \emph{IEEE Transactions on Smart Grid}, vol.~11, no.~4, pp.
  2929--2941, July 2020.

\bibitem{Siano_Systems_Sept_2019}
P.~{Siano}, G.~{De Marco}, A.~{Rol\'{a}n}, and V.~{Loia}, ``A survey and
  evaluation of the potentials of distributed ledger technology for
  peer-to-peer transactive energy exchanges in local energy markets,''
  \emph{IEEE Systems Journal}, vol.~13, no.~3, pp. 3454--3466, Sept. 2019.

\bibitem{Zhang_EP_May_2017}
C.~Zhang, J.~Wu, C.~Long, and M.~Cheng, ``Review of existing peer-to-peer
  energy trading projects,'' \emph{Energy Procedia}, vol. 105, pp. 2563--2568,
  May 2017.

\bibitem{Andoni_RSER_Feb_2019}
M.~Andoni, V.~Robu, D.~Flynn, S.~Abram, D.~Geach, D.~Jenkins, P.~McCallum, and
  A.~Peacock, ``{Blockchain technology in the energy sector: A systematic
  review of challenges and opportunities},'' \emph{Renewable and Sustainable
  Energy Reviews}, vol. 100, pp. 143--174, Feb. 2019.

\bibitem{Tushar_SPM_July_2018}
W.~{Tushar}, C.~{Yuen}, H.~{Mohsenian-Rad}, T.~{Saha}, H.~V. {Poor}, and K.~L.
  {Wood}, ``Transforming energy networks via peer-to-peer energy trading: The
  potential of game-theoretic approaches,'' \emph{IEEE Signal Processing
  Magazine}, vol.~35, no.~4, pp. 90--111, July 2018.

\bibitem{Juhar_Energies_June_2018}
J.~Abdella and K.~Shuaib, ``Peer to peer distributed energy trading in smart
  grids: {A} survey,'' \emph{MDPI Energies}, vol.~11, no.~6, pp. 1--22, June
  2018.

\bibitem{Sousa_RSER_Apr_2019}
T.~Sousa, T.~Soares, P.~Pinson, F.~Moret, T.~Baroche, and E.~Sorin,
  ``{Peer-to-peer and community-based markets: A comprehensive review},''
  \emph{Renewable and Sustainable Energy Reviews}, vol. 104, pp. 367--378, Apr.
  2019.

\bibitem{Guerrero_RSER_Oct_2020}
J.~Guerrero, D.~Gebbran, S.~Mhanna, A.~C. Chapman, and G.~Verbi\v{c}, ``Towards
  a transactive energy system for integration of distributed energy resources:
  Home energy management, distributed optimal power flow, and peer-to-peer
  energy trading,'' \emph{Renewable and Sustainable Energy Reviews}, vol. 132,
  pp. 110\,000:1--27, Oct. 2020.

\bibitem{RalphSims_BookChp_2007}
R.~E.~H. Sims, R.~N. Schock, A.~Adegbululgbe, J.~Fenhann,
  I.~Konstantinaviciute, W.~Moomaw, H.~B. Nimir, B.~Schlamadinger,
  J.~Torres-Mart\'{i}nez, C.~Turner, Y.~Uchiyama, S.~J. Vuori, N.~Wamukonya,
  and X.~Zhang, ``Energy supply,'' in \emph{Climate Change 2007: Mitigation of
  Climate Change}, B.~Metz, O.~Davidson, P.~Bosch, R.~Dave, and L.~Meyer,
  Eds.\hskip 1em plus 0.5em minus 0.4em\relax Cambridge, New York: Cambridge
  University Press, 2007, ch.~4, pp. 262--322.

\bibitem{Peck_IEEESpectrum_Oct_2017}
M.~E. {Peck} and D.~{Wagman}, ``Energy trading for fun and profit buy your
  neighbor's rooftop solar power or sell your own-it'll all be on a
  blockchain,'' \emph{IEEE Spectrum}, vol.~54, no.~10, pp. 56--61, Oct. 2017.

\bibitem{ThomasBauwens_March_2016}
T.~Bauwens, B.~Gotchev, and L.~Holstenkamp, ``{What drives the development of
  community energy in Europe? The case of wind power cooperatives},''
  \emph{Energy Research and Social Science}, vol.~13, pp. 136--147, Mar. 2016.

\bibitem{Abrishambaf_ESR_Nov_2019}
O.~Abrishambaf, F.~Lezama, P.~Faria, and Z.~Vale, ``{Towards transactive energy
  systems: An analysis on current trends},'' \emph{Energy Strategy Reviews},
  vol.~26, pp. 100\,418:1--17, Nov. 2019.

\bibitem{Mohsen_Energies_Apr_2020}
M.~Khorasany, D.~Azuatalam, R.~Glasgow, A.~Leibman, and R.~Razzaghi,
  ``Transactive energy market for energy management in microgrids: {The Monash}
  microgrid case study,'' \emph{MDPI Energies}, vol.~13, no.~8, pp. 2010:1--23,
  Apr. 2020.

\bibitem{Alam_AE_Mar_2019}
M.~R. Alam, M.~St-Hilaire, and T.~Kunz, ``{Peer-to-peer energy trading amont
  smart homes},'' \emph{Applied Energy}, vol. 238, pp. 1434--1443, Mar. 2019.

\bibitem{Tushar_NE_2020}
\BIBentryALTinterwordspacing
W.~Tushar, T.~K. Saha, C.~Yuen, D.~Smith, P.~Ashworth, H.~V. Poor, and
  S.~Basnet, ``{Challenges and prospects for negawatt trading in light of
  recent technological developments},'' \emph{Nature Energy}, Aug. 2020.
  [Online]. Available: \url{https://doi.org/10.1038/s41560-020-0671-0}
\BIBentrySTDinterwordspacing

\bibitem{Tushar_TSG_Mar_2020}
W.~{Tushar}, T.~K. {Saha}, C.~{Yuen}, T.~{Morstyn}, {Nahid-Al-Masood}, H.~V.
  {Poor}, and R.~{Bean}, ``Grid influenced peer-to-peer energy trading,''
  \emph{IEEE Transactions on Smart Grid}, vol.~11, no.~2, pp. 1407--1418, Mar.
  2020.

\bibitem{Esther_AE_Jan_2018}
E.~Mengelkamp, J.~G\"{a}rttner, K.~Rock, S.~Kessler, L.~Orsini, and
  C.~Weinhardt, ``{Designing microgrid energy markets - A case study: The
  Brooklyn Microgrid},'' \emph{Applied Energy}, vol. 210, pp. 870--880, Jan.
  2018.

\bibitem{Jogunola_Energies_Dec_2018}
O.~Jogunola, A.~Ikpehai, K.~Anoh, B.~Adebisi, M.~Hammoudeh, H.~Gacanin, and
  G.~Harris, ``Comparative analysis of {P2P} architecture for energy trading
  and sharing,'' \emph{MDPI Energies}, vol.~11, no.~1, pp. 62:1--62:20, Dec.
  2018.

\bibitem{Tushar_TSG_May_2016}
W.~{Tushar}, B.~{Chai}, C.~{Yuen}, S.~{Huang}, D.~B. {Smith}, H.~V. {Poor}, and
  Z.~{Yang}, ``Energy storage sharing in smart grid: A modified auction-based
  approach,'' \emph{IEEE Transactions on Smart Grid}, vol.~7, no.~3, pp.
  1462--1475, May 2016.

\bibitem{Zhou_Engineering_June_2020}
\BIBentryALTinterwordspacing
Y.~Zhou, J.~Wu, C.~Long, and W.~Ming, ``{State-of-the-art analysis and
  perspectives for peer-to-peer energy trading},'' \emph{Engineering}, June
  2020, pre-print. [Online]. Available:
  \url{https://doi.org/10.1016/j.eng.2020.06.002}
\BIBentrySTDinterwordspacing

\bibitem{Papadaskalopoulos_TPWRS_Nov_2013}
D.~{Papadaskalopoulos} and G.~{Strbac}, ``Decentralized participation of
  flexible demand in electricity markets?part i: Market mechanism,'' \emph{IEEE
  Transactions on Power Systems}, vol.~28, no.~4, pp. 3658--3666, Nov. 2013.

\bibitem{Hou_TII_June_2019}
W.~{Hou}, L.~{Guo}, and Z.~{Ning}, ``Local electricity storage for
  blockchain-based energy trading in industrial {Internet of Things},''
  \emph{IEEE Transactions on Industrial Informatics}, vol.~15, no.~6, pp.
  3610--3619, June 2019.

\bibitem{Thomas_TSG_Mar_2019}
T.~{Morstyn}, A.~{Teytelboym}, and M.~D. {Mcculloch}, ``Bilateral contract
  networks for peer-to-peer energy trading,'' \emph{IEEE Transactions on Smart
  Grid}, vol.~10, no.~2, pp. 2026--2035, Mar. 2019.

\bibitem{Sorin_TPWRS_Mar_2019}
E.~{Sorin}, L.~{Bobo}, and P.~{Pinson}, ``Consensus-based approach to
  peer-to-peer electricity markets with product differentiation,'' \emph{IEEE
  Transactions on Power Systems}, vol.~34, no.~2, pp. 994--1004, Mar. 2019.

\bibitem{Khorasany_TII_July_2020}
M.~{Khorasany}, Y.~{Mishra}, and G.~{Ledwich}, ``A decentralized bilateral
  energy trading system for peer-to-peer electricity markets,'' \emph{IEEE
  Transactions on Industrial Electronics}, vol.~67, no.~6, pp. 4646--4657, July
  2020.

\bibitem{Paudel_TIE_Aug_2019}
A.~{Paudel}, K.~{Chaudhari}, C.~{Long}, and H.~B. {Gooi}, ``Peer-to-peer energy
  trading in a prosumer-based community microgrid: A game-theoretic model,''
  \emph{IEEE Transactions on Industrial Electronics}, vol.~66, no.~8, pp.
  6087--6097, Aug. 2019.

\bibitem{Gonzalez_IEM_Dec_2018}
P.~{Baez-Gonzalez}, E.~{Rodriguez-Diaz}, J.~C. {Vasquez}, and J.~M. {Guerrero},
  ``Peer-to-peer energy market for community microgrids [technology leaders],''
  \emph{IEEE Electrification Magazine}, vol.~6, no.~4, pp. 102--107, Dec. 2018.

\bibitem{Moret_TPWRS_EA_2018}
F.~{Moret} and P.~{Pinson}, ``Energy collectives: {A} community and fairness
  based approach to future electricity markets,'' \emph{IEEE Transactions on
  Power Systems}, vol.~34, no.~5, pp. 3994--4004, Sept. 2019.

\bibitem{Zia_Access_Jan_2020}
M.~F. {Zia}, M.~{Benbouzid}, E.~{Elbouchikhi}, S.~M. {Muyeen}, K.~{Techato},
  and J.~M. {Guerrero}, ``Microgrid transactive energy: Review, architectures,
  distributed ledger technologies, and market analysis,'' \emph{IEEE Access},
  vol.~8, pp. 19\,410--19\,432, Jan. 2020.

\bibitem{Li_TII_Aug_2018}
Z.~{Li}, J.~{Kang}, R.~{Yu}, D.~{Ye}, Q.~{Deng}, and Y.~{Zhang}, ``Consortium
  blockchain for secure energy trading in industrial internet of things,''
  \emph{IEEE Transactions on Industrial Informatics}, vol.~14, no.~8, pp.
  3690--3700, Aug. 2018.

\bibitem{Hashgraph_TVT_2020}
V.~{Hassija}, V.~{Saxena}, V.~{Chamola}, and F.~{Richard Yu}, ``A parking slot
  allocation framework based on virtual voting and adaptive pricing
  algorithm,'' \emph{IEEE Transactions on Vehicular Technology}, vol.~69,
  no.~6, pp. 5945--5957, June 2020.

\bibitem{Holochain_2020}
K.~Wahlstrom, A.~Ul-haq, and O.~Burmeister, ``Privacy by design: {A} holochain
  exploration,'' \emph{Australasian Journal of Infomration Systems}, vol.~24,
  pp. 1--9, June 2020.

\bibitem{DAG_AppliedEnergy_2015}
Q.~Ji and Y.~Fan, ``Dynamic integration of world oil prices: A reinvestigation
  of globalisation vs. regionalisation,'' \emph{Applied Energy}, vol. 155, pp.
  171--180, Oct. 2015.

\bibitem{Hyperledger_IBM_2019}
F.~{Benhamouda}, S.~{Halevi}, and T.~{Halevi}, ``Supporting private data on
  hyperledger fabric with secure multiparty computation,'' \emph{IBM Journal of
  Research and Development}, vol.~63, no. 2/3, pp. 3:1--3:8, Mar.-May 2019.

\bibitem{Ethereum_Access_2019}
A.~{Pinna}, S.~{Ibba}, G.~{Baralla}, R.~{Tonelli}, and M.~{Marchesi}, ``A
  massive analysis of ethereum smart contracts empirical study and code
  metrics,'' \emph{IEEE Access}, vol.~7, pp. 78\,194--78\,213, June 2019.

\bibitem{Chen_TCS_July_2019}
J.~Chen and S.~Micali, ``{Algorand: A secure and efficient distributed
  ledger},'' \emph{Theoretical Computer Science}, vol. 777, pp. 155--183, July
  2019.

\bibitem{Tushar_SPM_Sept_2018}
W.~{Tushar}, N.~{Wijerathne}, W.~{Li}, C.~{Yuen}, H.~V. {Poor}, T.~K. {Saha},
  and K.~L. {Wood}, ``Internet of things for green building management:
  {D}isruptive innovations through low-cost sensor technology and artificial
  intelligence,'' \emph{IEEE Signal Processing Magazine}, vol.~35, no.~5, pp.
  100--110, Sept. 2018.

\bibitem{Ali_TCE_Nov_2017}
A.~R. {Al-Ali}, I.~A. {Zualkernan}, M.~{Rashid}, R.~{Gupta}, and M.~{Alikarar},
  ``A smart home energy management system using iot and big data analytics
  approach,'' \emph{IEEE Transactions on Consumer Electronics}, vol.~63, no.~4,
  pp. 426--434, Nov. 2017.

\bibitem{Nasim_AE_Aug_2018}
N.~Sahraei, E.~E. Looney, S.~M. Watson, I.~M. Peters, and T.~Buonassisi,
  ``Adaptive power consumption improves the reliability of solar-powered
  devices for internet of things,'' \emph{Applied Energy}, vol. 224, pp.
  322--329, Aug. 2018.

\bibitem{Ethan_AE_Apr_2019}
E.~Png, S.~Srinivasan, K.~Bekiroglu, J.~Chaoyang, R.~Su, and K.~Poolla, ``An
  internet of things upgrade for smart and scalable heating, ventilation and
  air-conditioning control in commercial buildings,'' \emph{Applied Energy},
  vol. 239, pp. 408--424, Apr. 2019.

\bibitem{Bedi_IOT_Apr_2018}
G.~{Bedi}, G.~K. {Venayagamoorthy}, R.~{Singh}, R.~R. {Brooks}, and K.~{Wang},
  ``Review of {Internet of Things (IoT)} in electric power and energy
  systems,'' \emph{IEEE Internet of Things Journal}, vol.~5, no.~2, pp.
  847--870, Apr. 2018.

\bibitem{Reka_RSER_Aug_2018}
S.~S. Reka and T.~Dragicevic, ``{Future effectual role of energy delivery: A
  comprehensive review of Internet of Things and smart grid},'' \emph{Renewable
  and Sustainable Energy Reviews}, vol.~91, pp. 90--108, Aug. 2018.

\bibitem{Sarvapali_ACM_Apr_2012}
S.~D. Ramchurn, P.~Vytelingum, A.~Rogers, and N.~R. Jennings, ``{Putting the
  `smarts' into the smart grid: A grand challenge for artificial
  intelligence},'' \emph{Communications of the ACM}, vol.~55, no.~4, pp.
  88--97, Apr. 2012.

\bibitem{Yuzchen_AE_Feb_2018}
Y.~Peng, A.~Rysanek, Z.~Nagy, and A.~Schl\"{u}ter, ``{Using machine learning
  techniques for occupancy-prediction-based cooling control in office
  buildings},'' \emph{Applied Energy}, vol. 211, pp. 1343--1358, Feb. 2018.

\bibitem{Jose_AE_Feb_2019}
J.~R. V\'{a}zquez-Canteli and Z.~Nagy, ``{Reinforcement learning for demand
  response: A review of algorithms and modeling techniques},'' \emph{Applied
  Energy}, vol. 235, pp. 1072--1089, Feb. 2019.

\bibitem{Konstantakopoulos_AE_Mar_2019}
I.~C. Konstantakopoulos, A.~R. arkan, S.~He, T.~Veeravalli, H.~Liu, and
  C.~Spanos, ``{A deep learning and gamification approach to improving
  human-building interaction and energy efficiency in smart infrastructure},''
  \emph{Applied Energy}, vol. 237, pp. 810--821, Mar. 2019.

\bibitem{Reynold_AE_May_2018}
J.~Reynolds, Y.~Rezgui, A.~Kwan, and S.~Piriou, ``{A zone-level, building
  energy optimisation combining an artificial neural network, a genetic
  algorithm, and model predictive control},'' \emph{Energy}, vol. 151, pp.
  729--739, May 2018.

\bibitem{Perry_GEB_Oct_2019}
\BIBentryALTinterwordspacing
C.~Perry, H.~Bastian, , and D.~York, ``{Grid-interactive efficient building
  utility programs: State of the market},'' American Council for an
  Energy-Efficient Economy, Washington, DC, Tech. Rep., Oct. 2019. [Online].
  Available: \url{https://www.aceee.org/sites/default/files/gebs-103019.pdf}
\BIBentrySTDinterwordspacing

\bibitem{Neukomm_GEB_Apr_2019}
\BIBentryALTinterwordspacing
M.~Neukomm, V.~Nubbe, and R.~Fares, ``{Grid-interactive efficient buildings -
  Overview},'' Office of Energy Efficiency and Renewable Energy, U.S.
  Department of Energy, Washington, DC, Tech. Rep., Apr. 2019. [Online].
  Available:
  \url{https://www.energy.gov/sites/prod/files/2019/04/f61/bto-geb_overview-4.15.19.pdf}
\BIBentrySTDinterwordspacing

\bibitem{Arturo_PhDThesis_2009}
A.~D. Alarc\'{o}n-Rodr\'{i}guez, ``A multi-objective planning framework for
  analysing the integration of distributed energy resources,'' {PhD}
  dissertation, Department of Electronic and Electrical Engineering, University
  of Strathclyde, Glasgow, Scotland, UK, Apr. 2009.

\bibitem{Yan_TPWRS_Sept_2018}
R.~{Yan}, N.~{-Masood}, T.~{Kumar Saha}, F.~{Bai}, and H.~{Gu}, ``The anatomy
  of the 2016 south australia blackout: A catastrophic event in a high
  renewable network,'' \emph{IEEE Transactions on Power Systems}, vol.~33,
  no.~5, pp. 5374--5388, Sept. 2018.

\bibitem{Weckx_TSE_Oct_2016}
S.~{Weckx} and J.~{Driesen}, ``Optimal local reactive power control by pv
  inverters,'' \emph{IEEE Transactions on Sustainable Energy}, vol.~7, no.~4,
  pp. 1624--1633, Oct. 2016.

\bibitem{Jahedul_RSER_Oct_2020}
J.~I. Chowdhury, N.~Balta-Ozkan, P.~Goglio, Y.~Hu, L.~Varga, and L.~McCabe,
  ``Techno-environmental analysis of battery storage for grid level energy
  services,'' \emph{Renewable and Sustainable Energy Reviews}, vol. 131, p.
  110018, Oct. 2020.

\bibitem{Li_TSG_July_2019}
T.~{Li} and M.~{Dong}, ``Residential energy storage management with
  bidirectional energy control,'' \emph{IEEE Transactions on Smart Grid},
  vol.~10, no.~4, pp. 3596--3611, July 2019.

\bibitem{Kikusato_TSG_May_2019}
H.~{Kikusato}, K.~{Mori}, S.~{Yoshizawa}, Y.~{Fujimoto}, H.~{Asano},
  Y.~{Hayashi}, A.~{Kawashima}, S.~{Inagaki}, and T.~{Suzuki}, ``Electric
  vehicle charge?discharge management for utilization of photovoltaic by
  coordination between home and grid energy management systems,'' \emph{IEEE
  Transactions on Smart Grid}, vol.~10, no.~3, pp. 3186--3197, May 2019.

\bibitem{Kang_TII_Dec_2017}
J.~{Kang}, R.~{Yu}, X.~{Huang}, S.~{Maharjan}, Y.~{Zhang}, and E.~{Hossain},
  ``Enabling localized peer-to-peer electricity trading among plug-in hybrid
  electric vehicles using consortium blockchains,'' \emph{IEEE Transactions on
  Industrial Informatics}, vol.~13, no.~6, pp. 3154--3164, Dec. 2017.

\bibitem{Kristina_Thesis_2012}
\BIBentryALTinterwordspacing
K.~Hojckova, ``Watt's next: {On} socio-technical transition towards future
  electricity system architectures,'' {PhD} dissertation, Chalmers University
  of Technology, Gothenburg, Sweden, 2012. [Online]. Available:
  \url{https://research.chalmers.se/publication/505308/file/505308_Fulltext.pdf}
\BIBentrySTDinterwordspacing

\bibitem{Tushar_Energy_Apr_2020}
W.~Tushar, L.~Lan, C.~Withanage, H.~E.~K. Sng, C.~Yuen, K.~L. Wood, and T.~K.
  Saha, ``{Exploiting design thinking to improve energy efficiency of
  buildings},'' \emph{Energy}, vol. 197, pp. 117\,141:1--16, Apr. 2020.

\bibitem{Sachs_ProceedingIEEE_Feb_2019}
J.~{Sachs}, L.~A.~A. {Andersson}, J.~{Ara{\`u}jo}, C.~{Curescu},
  J.~{Lundsj{\"o}}, G.~{Rune}, E.~{Steinbach}, and G.~{Wikstr{\"o}m},
  ``Adaptive 5g low-latency communication for tactile internet services,''
  \emph{Proceedings of the IEEE}, vol. 107, no.~2, pp. 325--349, Feb. 2019.

\bibitem{Manic_IEM_Mar_2016}
M.~{Manic}, D.~{Wijayasekara}, K.~{Amarasinghe}, and J.~J. {Rodriguez-Andina},
  ``Building energy management systems: The age of intelligent and adaptive
  buildings,'' \emph{IEEE Industrial Electronics Magazine}, vol.~10, no.~1, pp.
  25--39, Mar. 2016.

\bibitem{RuiJing_AE_Mar_2020}
R.~Jing, M.~N. Xie, F.~X. Wang, and L.~X. Chen, ``{Fair P2P energy trading
  between residential and commercial multi-energy systems enabling integrated
  demand-side management},'' \emph{Applied Energy}, vol. 262, pp.
  114\,551:1--17, Mar. 2020.

\bibitem{Alam_EE_Dec_2017}
M.~R. Alam, M.~St-Hilaire, and T.~Kunz, ``{An optimal P2P energy trading model
  for smart homes in the smart grid},'' \emph{Energy Efficiency}, vol.~10, pp.
  1475--1493, Dec. 2017.

\bibitem{Ippolito_ENB_Feb_2014}
M.~G. Ippolito, E.~R. Sanseverino, and G.~Zizzo, ``{Impact of building
  automation control systems and technical building management systems on the
  energy performance class of residential buildings: An Italian case study},''
  \emph{Energy and Buildings}, vol.~69, pp. 33--40, Feb. 2014.

\bibitem{Lazos_RSER_Nov_2014}
D.~Lazos, A.~B. Sproul, and M.~Kay, ``{Optimisation of energy management in
  commercial buildings with weather forecasting inputs: A review},''
  \emph{Renewable and Sustainable Energy Reviews}, vol.~39, pp. 587--603, Nov.
  2014.

\bibitem{WenTai_TETC_Sept_2019}
W.~{Li}, S.~R. {Gubba}, W.~{Tushar}, C.~{Yuen}, N.~U. {Hassan}, H.~V. {Poor},
  K.~L. {Wood}, and C.~{Wen}, ``Data driven electricity management for
  residential air conditioning systems: An experimental approach,'' \emph{IEEE
  Transactions on Emerging Topics in Computing}, vol.~7, no.~3, pp. 380--391,
  July-Sept. 2019.

\bibitem{Marco_AE_June_2020}
M.~Beccali, L.~Bellia, F.~Fragliasso, M.~Bonomolo, G.~Zizzo, and G.~Spada,
  ``{Assessing the lighting systems flexibility for reducing and managing the
  power peaks in smart grids},'' \emph{Applied Energy}, vol. 268, pp.
  114\,924:1--16, June 2020.

\bibitem{PDu_TSG_June_2011}
P.~{Du} and N.~{Lu}, ``Appliance commitment for household load scheduling,''
  \emph{IEEE Transactions on Smart Grid}, vol.~2, no.~2, pp. 411--419, June
  2011.

\bibitem{Raja_TETCI_2020}
\BIBentryALTinterwordspacing
B.~{Rajasekhar}, W.~{Tushar}, C.~{Lork}, Y.~{Zhou}, C.~{Yuen}, N.~M.
  {Pindoriya}, and K.~L. {Wood}, ``A survey of computational intelligence
  techniques for air-conditioners energy management,'' \emph{IEEE Transactions
  on Emerging Topics in Computational Intelligence}, pp. 1--16, 2020, early
  access. [Online]. Available: \url{https://doi.org/10.1109/TETCI.2020.2991728}
\BIBentrySTDinterwordspacing

\bibitem{ASHARE_2006}
N.~Lu and S.~Katipamula, ``Evaluation of residential hvac control strategies
  for demand response programs,'' \emph{ASHRAE Transactions}, vol. 112, no.~1,
  pp. 535:1--12, 2006.

\bibitem{Gu_Ashrae_2012}
L.~Gu and R.~Raustad, ``{Short-term curtailment of HVAC loads in buildings},''
  \emph{ASHRAE Transactions}, vol. 118, pp. 467--474, 2012.

\bibitem{Yang_ENB_Aug_2014}
Z.~Yang and B.~Becerik-Gerber, ``{The coupled effects of personalized occupancy
  profile based HVAC schedules and room reassignment on building energy use},''
  \emph{Energy and Buildings}, vol.~78, pp. 113--122, Aug. 2014.

\bibitem{Su_IFAC_2014}
Y.~Su, R.~Su, and K.~Poolla, ``{Distributed Scheduling for Efficient HVAC
  Pre-cooling Operations},'' \emph{IFAC Proceedings Volumes}, vol.~47, no.~3,
  pp. 10\,451--10\,456, 2014.

\bibitem{Gayeski_HVAC_Sept_2012}
N.~T. Gayeski, P.~R. Armstrong, and L.~K. Norford, ``{Predictive pre-cooling of
  thermo-active building systems with low-lift chillers},'' \emph{HVAC\&R
  Research}, vol.~18, no.~5, pp. 858--873, Sept. 2012.

\bibitem{Nikdel_BNE_Feb_2018}
L.~Nikdel, K.~Janoyan, S.~D. Bird, and S.~E. Powers, ``{Multiple perspectives
  of the value of occupancy-based HVAC control systems},'' \emph{Building and
  Environment}, vol. 129, pp. 15--25, Feb. 2018.

\bibitem{Ngarambe_ENB_Mar_2020}
J.~Ngarambe, G.~Y. Yun, and M.~Santamouris, ``{The use of artificial
  intelligence (AI) methods in the prediction of thermal comfort in buildings:
  Energy implications of AI-based thermal comfort controls},'' \emph{Energy and
  Buildings}, vol. 211, pp. 109\,807:1--15, Mar. 2020.

\bibitem{Lork_AE_Oct_2020}
C.~Lork, W.-T. Li, Y.~Qin, Y.~Zhou, C.~Yuen, W.~Tushar, and T.~K. Saha, ``{An
  uncertainty-aware deep reinforcement learning framework for residential air
  conditioning energy management},'' \emph{Applied Energy}, vol. 276, pp.
  115\,426:1--12, Oct. 2020.

\bibitem{Zahra_EPSR_Nov_2020}
Z.~Rahimpour, G.~Verbi\v{c}, and A.~C. Chapman, ``Actor-critic learning for
  optimal building energy management with phase change materials,''
  \emph{Electric Power System Research}, vol. 188, pp. 106\,543:1--7, Nov.
  2020.

\bibitem{Valladares_BNE_May_2019}
W.~Valladares, M.~Galindo, J.~Guti\'{e}rrez, W.-C. Wu, K.-K. Liao, J.-C. Liao,
  K.-C. Lu, and C.-C. Wang, ``{Energy optimization associated with thermal
  comfort and indoor air control via a deep reinforcement learning
  algorithm},'' \emph{Building and Environment}, vol. 155, pp. 105--117, May
  2019.

\bibitem{Nagy_ENB_May_2015}
Z.~Nagy, F.~Y. Yong, M.~Frei, and A.~Schlueter, ``{Occupant centered lighting
  control for comfort and energy efficient building operation},'' \emph{Energy
  and Buildings}, vol.~94, pp. 100--108, May 2015.

\bibitem{Bakker_BNE_Feb_2017}
C.~de~Bakker, M.~Aries, H.~Kort, and A.~Rosemann, ``{Occupancy-based lighting
  control in open-plan office spaces: A state-of-the-art review},''
  \emph{Building and Environment}, vol. 112, pp. 308--321, Feb. 2017.

\bibitem{Oldewurtel_AE_Jan_2013}
F.~Oldewurtel, D.~Sturzenegger, and M.~Morari, ``{Importance of occupancy
  information for building climate control},'' \emph{Applied Energy}, vol. 101,
  pp. 521--532, Jan. 2013.

\bibitem{Park_BNE_Jan_2019}
J.~Y. Park, T.~Dougherty, H.~Fritz, and Z.~Nagy, ``{LightLearn: An adaptive and
  occupant centered controller for lighting based on reinforcement learning},''
  \emph{Building and Environment}, vol. 147, pp. 397--414, Jan. 2019.

\bibitem{Zou_ENB_Jan_2018}
H.~Zou, Y.~Zhou, H.~Jing, S.-C. Chien, L.~Xie, and C.~J. Spanos, ``{WinLight: A
  WiFi-based occupancy-driven lighting control system for smart building},''
  \emph{Energy and Buildings}, vol. 158, pp. 924--938, Jan. 2018.

\bibitem{Labeodan_ENB_Sept_2016}
T.~Labeodan, C.~D. Bakker, A.~Rosemann, and W.~Zeiler, ``{On the application of
  wireless sensors and actuators network in existing buildings for occupancy
  detection and occupancy-driven lighting control},'' \emph{Energy and
  Buildings}, vol. 127, pp. 75--83, Sept. 2016.

\bibitem{Meugheuvel_ENB_June_2014}
N.~V. de~Meugheuvel, A.~Pandharipande, D.~Caicedo, and P.~P.~J. van~den Hof,
  ``{Distributed lighting control with daylight and occupancy adaptation},''
  \emph{Energy and Buildings}, vol.~75, pp. 321--329, June 2014.

\bibitem{Liu_ENB_Sept_2016}
J.~Liu, W.~Zhang, X.~Chu, and Y.~Liu, ``{Fuzzy logic controller for energy
  savings in a smart LED lighting system considering lighting comfort and
  daylight},'' \emph{Energy and Buildings}, vol. 127, pp. 95--104, Sept. 2016.

\bibitem{Haq_RSER_May_2014}
M.~A.~U. Haq, M.~Y. Hassan, H.~Abdullah, H.~A. Rahman, M.~P. Abdullah, H.~F,
  and D.~M. Said, ``{A review on lighting control technologies in commercial
  buildings, their performance and affecting factors},'' \emph{Renewable and
  Sustainable Energy Reviews}, vol.~33, pp. 268--279, May 2014.

\bibitem{Li_CSEE_Dec_2018}
R.~{Li} and S.~{You}, ``Exploring potential of energy flexibility in buildings
  for energy system services,'' \emph{CSEE Journal of Power and Energy
  Systems}, vol.~4, no.~4, pp. 434--443, Dec. 2018.

\bibitem{Georges_AE_Feb_2017}
E.~Georges, B.~Corn\'{e}lusse, D.~Ernst, V.~Lemort, and S.~Mathlieu,
  ``{Residential heat pump as flexible load for direct control service with
  parametrized duration and rebound effect},'' \emph{Applied Energy}, vol. 187,
  pp. 140--153, Feb. 2017.

\bibitem{NaveedHassan_TSG_Nov_2015}
N.~{Ul Hassan}, Y.~I. {Khalid}, C.~{Yuen}, and W.~{Tushar}, ``Customer
  engagement plans for peak load reduction in residential smart grids,''
  \emph{IEEE Transactions on Smart Grid}, vol.~6, no.~6, pp. 3029--3041, Nov.
  2015.

\bibitem{Huang_TPWRS_Aug_2004}
{Kun-Yuan Huang} and {Yann-Chang Huang}, ``Integrating direct load control with
  interruptible load management to provide instantaneous reserves for ancillary
  services,'' \emph{IEEE Transactions on Power Systems}, vol.~19, no.~3, pp.
  1626--1634, Aug. 2004.

\bibitem{Bhattacharya_TPWRS_May_2000}
K.~{Bhattacharya}, M.~H.~J. {Bollen}, and J.~E. {Daalder}, ``Real time optimal
  interruptible tariff mechanism incorporating utility-customer interactions,''
  \emph{IEEE Transactions on Power Systems}, vol.~15, no.~2, pp. 700--706, May
  2000.

\bibitem{Habib_TEES_Nov_2016}
H.~A. Aalami and A.~Khatibzadeh, ``Regulation of market clearing price based on
  nonlinear models of demand bidding and emergency demand response programs,''
  \emph{International Transactions on Electrical Energy Systems}, vol.~26,
  no.~11, pp. 2463--2478, Nov. 2016.

\bibitem{Wang_TSG_July_2020}
B.~{Wang}, Y.~{Li}, W.~{Ming}, and S.~{Wang}, ``Deep reinforcement learning
  method for demand response management of interruptible load,'' \emph{IEEE
  Transactions on Smart Grid}, vol.~11, no.~4, pp. 3146--3155, July 2020.

\bibitem{Andersen_AE_Dec_2017}
F.~M. Andersen, M.~Baldini, L.~G. Hansen, and C.~L. Jensen, ``Households'
  hourly electricity consumption and peak demand in denmark,'' \emph{Applied
  Energy}, vol. 208, pp. 607--619, Dec. 2017.

\bibitem{Hussain_EJ_June_2018}
M.~Hussain and Y.~Gao, ``A review of demand response in an efficient smart grid
  environment,'' \emph{The Electricity Journal}, vol.~31, pp. 55--63, June
  2018.

\bibitem{Wakui_Energy_Feb_2010}
T.~Wakui, R.~Yokoyama, and K.~ichi Shimizu, ``{Suitable operational strategy
  for power interchange operation using multiple residential SOFC (solid oxide
  fuel cell) cogeneration systems},'' \emph{Energy}, vol.~35, no.~2, pp.
  740--750, Feb. 2010.

\bibitem{Aki_Hydrogen_Nov_2016}
H.~Aki, T.~Wakui, and R.~Yokoyama, ``{Development of an energy management
  system for optimal operation of fuel cell based residential energy
  systems},'' \emph{International Journal of Hydrogen Energy}, vol.~41, no.~44,
  pp. 20\,314--20\,325, Nov. 2016.

\bibitem{Aki_Energy_June_2018}
H.~Aki, T.~Wakui, R.~Yokoyama, and K.~Sawada, ``{Optimal management of multiple
  heat sources in a residential area by an energy management system},''
  \emph{Energy}, vol. 153, pp. 1048--1060, June 2018.

\bibitem{Tran_ECM_Oct_2018}
H.~N. Tran, T.~Narikiyo, M.~Kawanishi, S.~Kikuchi, and S.~Takaba, ``{Whole-day
  optimal operation of multiple combined heat and power systems by alternating
  direction method of multipliers and consensus theory},'' \emph{Energy
  Conversion and Management}, vol. 174, pp. 475--488, Oct. 2018.

\bibitem{Wayes_AE_Mar_2020}
W.~Tushar, T.~K. Saha, C.~Yuen, M.~I. Azim, T.~Morstyn, H.~V. Poor, D.~Niyato,
  and R.~Bean, ``{A coalition formation game framework for peer-to-peer energy
  trading},'' \emph{Applied Energy}, vol. 261, pp. 114\,436:1--13, Mar. 2020.

\bibitem{Jazizadeh_AE_June_2018}
F.~Jazizadeh and W.~Jung, ``{Personalized thermal comfort inference using RGB
  video images for distributed HVAC control},'' \emph{Applied Energy}, vol.
  220, pp. 829--841, June 2018.

\bibitem{Jung_BE_July_2019}
W.~Jung and F.~Jazizadeh, ``{Comparative assessment of HVAC control strategies
  using personal thermal comfort and sensitivity models},'' \emph{Building and
  Environment}, vol. 158, pp. 104--119, July 2019.

\bibitem{Adhikari_TSG_May_2020}
R.~{Adhikari}, M.~{Pipattanasomporn}, and S.~{Rahman}, ``Heuristic algorithms
  for aggregated hvac control via smart thermostats for regulation service,''
  \emph{IEEE Transactions on Smart Grid}, vol.~11, no.~3, pp. 2023--2032, May
  2020.

\bibitem{Jiang_EPSR_Sept_2020}
X.~Jiang and L.~Wu, ``A residential load scheduling based on cost efficiency
  and consumer's preference for demand response in smart grid,'' \emph{IEEE
  Transactions on Smart Grid}, vol. 186, pp. 106\,410:1--10, Sept. 2020.

\bibitem{Tushar_Access_Oct_2018}
W.~{Tushar}, T.~K. {Saha}, C.~{Yuen}, P.~{Liddell}, R.~{Bean}, and H.~V.
  {Poor}, ``Peer-to-peer energy trading with sustainable user participation:
  {A} game theoretic approach,'' \emph{IEEE Access}, vol.~6, pp.
  62\,932--62\,943, Oct. 2018.

\bibitem{Wilkinson_ERSC_Aug_2020}
S.~Wilkinson, K.~Hojckova, C.~Eon, G.~M. Norrison, and B.~Sand\'{e}n, ``{Is
  peer-to-peer electricity trading empowering users? Evidence on motivations
  and roles in a prosumer business model trial in Australia},'' \emph{Energy
  Research \& Social Science}, vol.~66, pp. 101\,500:1--23, Aug. 2020.

\bibitem{Nguyen_AE_Oct_2018}
S.~Nguyen, W.~Peng, P.~Sokolowski, D.~Alahakoon, and X.~Yu, ``{Optimizing
  rooftop photovoltaic distributed generation with battery storage for
  peer-to-peer energy trading},'' \emph{Applied Energy}, vol. 228, pp.
  2567--2580, Oct. 2018.

\bibitem{Long_AE_Sep_2018}
C.~Long, J.~Wu, Y.~Zhou, and N.~Jenkins, ``{Peer-to-peer energy sharing through
  a two-stage aggregated battery control in a community Microgrid},''
  \emph{Applied Energy}, vol. 226, pp. 261--276, Sep. 2018.

\bibitem{Jan_EB_Feb_2019}
J.~M. Zepter, A.~L\"{u}th, P.~C. del Granado, and R.~Egging, ``Prosumer
  integration in wholesale electricity markets: {S}ynergies of peer-to-peer
  trade and residential storage,'' \emph{Energy and Buildings}, vol. 184, pp.
  163--176, Feb. 2019.

\bibitem{Nizami_AE_Mar_2020}
M.~S.~H. Nizami, M.~J. Hossain, B.~M.~R. Amin, and E.~Fernandez, ``A
  residential energy management system with bi-level optimization-based bidding
  strategy for day-ahead bi-directional electricity trading,'' \emph{Applied
  Energy}, vol. 261, pp. 114\,322:1--17, Mar. 2020.

\bibitem{Shantonu_AE_Nov_2018}
S.~Chakraborty, T.~Baarslag, and M.~Kaisers, ``{Automated peer-to-peer
  negotiation for energy contract settlements in residential cooperatives},''
  \emph{Applied Energy}, vol. 259, pp. 114\,173:1--14, Feb. 2020.

\bibitem{Hangyue_IJEPES_July_2019}
H.~Liu, D.~Azuatalam, A.~C. Chapman, and G.~Verbi\v{c}, ``Techno-economic
  feasibility assessment of grid-defection,'' \emph{International Journal of
  Electrical Power \& Energy Systems}, vol. 10*, pp. 403--412, July 2019.

\bibitem{Markus_IJER_Feb_2020}
M.~F\"{o}rstl, D.~Azuatalam, A.~C. Chapman, G.~Verbi\v{c}, A.~Jossen, and
  H.~Hesse, ``Assessment of residential battery storage systems and operation
  strategies considering battery aging,'' \emph{International Journal of Energy
  Research}, vol.~44, no.~2, pp. 718--731, Feb. 2020.

\bibitem{Rodrigues_AE_Mar_2020}
D.~L. Rodrigues, X.~Ye, X.~Xia, and B.~Zhu, ``{Battery energy storage sizing
  optimisation for different ownership structures in a peer-to-peer energy
  sharing community},'' \emph{Applied Energy}, vol. 262, pp. 114\,498:1--11,
  Mar. 2020.

\bibitem{Guerrero_PESGM_2019}
J.~{Guerrero}, A.~C. {Chapman}, and G.~{Verbi\v{c}}, ``Trading arrangements and
  cost allocation in p2p energy markets on low-voltage networks,'' in
  \emph{IEEE Power Energy Society General Meeting (PESGM)}, Atlanta, GA, Aug.
  2019, pp. 1--5.

\bibitem{Wang_AE_Oct_2019}
J.~Wang, H.~Zhong, C.~Wu, E.~Du, Q.~Xia, and C.~Kang, ``{Incentivizing
  distributed energy resource aggregation in energy and capacity markets: An
  energy sharing scheme and mechanism design},'' \emph{Applied Energy}, vol.
  252, pp. 113\,741:1--113\,741:13, Oct. 2019.

\bibitem{Kirchhoff_AE_June_2019}
H.~Kirchhoff and K.~Strunz, ``{Key drivers for successful development of
  peer-to-peer microgrids for swarm electrification},'' \emph{Applied Energy},
  vol. 244, pp. 46--62, June 2019.

\bibitem{Li_AE_Aug_2019}
Y.~Li, W.~Yang, P.~He, C.~Chen, and X.~Wang, ``{Design and management of a
  distributed hybrid energy system through smart contract and blockchain},''
  \emph{Applied Energy}, vol. 248, pp. 390--405, Aug. 2019.

\bibitem{Koirala_AE_Dec_2018}
B.~P. Koirala, E.~Oost, and H.~der Windt, ``{Community energy storage: A
  responsible innovation towards a sustainable energy system?}'' \emph{Applied
  Energy}, vol. 231, pp. 570--585, Dec. 2018.

\bibitem{Ma_AUPEC_2017}
Y.~{Ma}, M.~S.~S. {Abad}, D.~{Azuatalam}, G.~{Verbi\v{c}}, and A.~{Chapman},
  ``Impacts of community and distributed energy storage systems on unbalanced
  low voltage networks,'' in \emph{Australasian Universities Power Engineering
  Conference (AUPEC)}, Melbourne, Australia, Nov. 2017, pp. 1--6.

\bibitem{Ma_PowerTech_2019}
Y.~{Ma}, G.~{Verbi\v{c}}, and A.~C. {Chapman}, ``Estimating the option value of
  grid-scale battery systems to distribution network service providers,'' in
  \emph{IEEE Milan PowerTech}, Milan, Italy, June 2019, pp. 1--6.

\bibitem{Barbour_AE_Feb_2018}
E.~Barbour, D.~Parra, Z.~Awwad, and M.~C.Gonz\`{a}lez, ``{Community energy
  storage: A smart choice for the smart grid?}'' \emph{Applied Energy}, vol.
  212, pp. 489--497, Feb. 2018.

\bibitem{Scheller_AE_July_2020}
F.~Scheller, R.~Burkhardt, R.~Schwarzeit, R.~McKenna, and T.~Bruckner,
  ``{Competition between simultaneous demand-side flexibility options: the case
  of community electricity storage systems},'' \emph{Applied Energy}, vol. 269,
  pp. 114\,969:1--16, July 2020.

\bibitem{Dong_RSER_Oct_2020}
S.~Dong, E.~Kremers, M.~Brucoli, R.~Rothman, and S.~Brown, ``{Improving the
  feasibility of household and community energy storage: A
  techno-enviro-economic study for the UK},'' \emph{Renewable and Sustainable
  Energy Reviews}, vol. 131, pp. 110\,009:1--17, Oct. 2020.

\bibitem{Tarek_ENB_Mar_2017}
T.~A. Skaif, A.~C. Luna, M.~G. Zapata, J.~M. Guerrero, and B.~Bellalta,
  ``{Reputation-based joint scheduling of households appliances and storage in
  a microgrid with a shared battery},'' \emph{Energy and Buildings}, vol. 138,
  pp. 228--239, Mar. 2017.

\bibitem{Hafiz_AE_Feb_2019}
F.~Hafiz, A.~R. de~Queiroz, P.~Fajri, and I.~Husain, ``{Energy management and
  optimal storage sizing for a shared community: A multi-stage stochastic
  programming approach},'' \emph{Applied Energy}, vol. 236, pp. 42--54, Feb.
  2019.

\bibitem{Kasmaei_TPWRS_Jan_2020}
M.~{Pourakbari-Kasmaei}, M.~{Asensio}, M.~{Lehtonen}, and J.~{Contreras},
  ``Trilateral planning model for integrated community energy systems and
  pv-based prosumers - {A} bilevel stochastic programming approach,''
  \emph{IEEE Transactions on Power Systems}, vol.~35, no.~1, pp. 346--361, Jan.
  2020.

\bibitem{Zhong_TSG_EAccess_2020}
\BIBentryALTinterwordspacing
W.~{Zhong}, K.~{Xie}, Y.~{Liu}, C.~{Yang}, and S.~{Xie}, ``Multi-resource
  allocation of shared energy storage: A distributed combinatorial auction
  approach,'' \emph{IEEE Transactions on Smart Grid}, 2020, early access.
  [Online]. Available: \url{https://doi.org/10.1109/TSG.2020.2986468}
\BIBentrySTDinterwordspacing

\bibitem{Kalathil_TSG_Jan_2019}
D.~{Kalathil}, C.~{Wu}, K.~{Poolla}, and P.~{Varaiya}, ``The sharing economy
  for the electricity storage,'' \emph{IEEE Transactions on Smart Grid},
  vol.~10, no.~1, pp. 556--567, Jan. 2019.

\bibitem{Siang_RSER_Apr_2013}
S.~F. Tie and C.~W. Tan, ``A review of energy sources and energy management
  system in electric vehicles,'' \emph{Renewable and Sustainable Energy
  Reviews}, vol.~20, pp. 82--102, Apr. 2013.

\bibitem{Mwasilu_RSER_Apr_2014}
F.~Mwasilu, J.~J. Justo, E.-K. Kim, T.~D. Do, and J.-W. Jung, ``Electric
  vehicles and smart grid interaction: A review on vehicle to grid and
  renewable energy sources integration,'' \emph{Renewable and Sustainable
  Energy Reviews}, vol.~34, pp. 501--516, June 2014.

\bibitem{Rahmani-Andebili_IET_Apr_2019}
M.~{Rahmani-Andebili}, ``{Vehicle-for-grid (VfG): A} mobile energy storage in
  smart grid,'' \emph{IET Generation, Transmission Distribution}, vol.~13,
  no.~8, pp. 1358--1368, Apr. 2019.

\bibitem{Zhou_ECM_Nov_2019}
Y.~Zhou and S.~Cao, ``{Energy flexibility investigation of advanced
  grid-responsive energy control strategies with the static battery and
  electric vehicles: A case study of a high-rise office building in Hong
  Kong},'' \emph{Energy Conversation and Management}, vol. 199, pp.
  111\,888:1--22, Nov. 2019.

\bibitem{Peng_RSER_Feb_2017}
C.~Peng, J.~Zou, and L.~Lian, ``{Dispatching strategies of electric vehicles
  participating in frequency regulation on power grid: A review},''
  \emph{Renewable and Sustainable Energy Reviews}, vol.~68, pp. 147--152, Feb.
  2017.

\bibitem{Alvaro-Hermana_TSM_Fall_2016}
R.~{Alvaro-Hermana}, J.~{Fraile-Ardanuy}, P.~J. {Zufiria}, L.~{Knapen}, and
  D.~{Janssens}, ``Peer to peer energy trading with electric vehicles,''
  \emph{IEEE Intelligent Transportation Systems Magazine}, vol.~8, no.~3, pp.
  33--44, Fall 2016.

\bibitem{Zhang_TITS_Jan_2019}
R.~{Zhang}, X.~{Cheng}, and L.~{Yang}, ``Flexible energy management protocol
  for cooperative ev-to-ev charging,'' \emph{IEEE Transactions on Intelligent
  Transportation Systems}, vol.~20, no.~1, pp. 172--184, Jan. 2019.

\bibitem{Dai_WirelessCommunications_June_2019}
Y.~{Dai}, D.~{Xu}, S.~{Maharjan}, G.~{Qiao}, and Y.~{Zhang}, ``Artificial
  intelligence empowered edge computing and caching for internet of vehicles,''
  \emph{IEEE Wireless Communications}, vol.~26, no.~3, pp. 12--18, June 2019.

\bibitem{Liu_Access_vol7_2019}
H.~{Liu}, Y.~{Zhang}, S.~{Zheng}, and Y.~{Li}, ``Electric vehicle power trading
  mechanism based on blockchain and smart contract in {V2G} network,''
  \emph{IEEE Access}, vol.~7, pp. 160\,546--160\,558, 2019.

\bibitem{Liu_WirelessCommunication_Feb_2019}
\BIBentryALTinterwordspacing
C.~Liu, K.~K. Chai, X.~Zhang, and Y.~Chen, ``{Peer-to-peer electricity trading
  system: smart contracts based proof-of-benefit consensus protocol},''
  \emph{Wireless Networks}, Feb. 2019. [Online]. Available:
  \url{https://doi.org/10.1007/s11276-019-01949-0}
\BIBentrySTDinterwordspacing

\bibitem{Ridoy_AE_Jan_2020}
R.~Das, Y.~Wang, G.~Putrus, R.~Kotter, M.~Marzband, B.~Herteleer, and
  J.~Warmerdam, ``{Multi-objective techno-economic-environmental optimisation
  of electric vehicle for energy services},'' \emph{Applied Energy}, vol. 257,
  pp. 113\,965:1--18, Jan. 2020.

\bibitem{Aznavi_TIA_EA_2020}
\BIBentryALTinterwordspacing
S.~{Aznavi}, P.~{Fajri}, M.~B. {Shadmand}, and A.~{Khoshkbar-Sadigh},
  ``Peer-to-peer operation strategy of {PV} equipped office buildings and
  charging stations considering electric vehicle energy pricing,'' \emph{IEEE
  Transactions on Industry Applications}, 2020, early access. [Online].
  Available: \url{https://doi.org/10.1109/TIA.2020.2990585}
\BIBentrySTDinterwordspacing

\bibitem{Stelmach_EP_Sept_2020}
G.~Stelmach, C.~Zanocco, J.~Flora, R.~Rajagopal, and H.~S. Boudet, ``Exploring
  household energy rules and activities during peak demand to better determine
  potential responsiveness to time-of-use pricing,'' \emph{Energy Policy}, vol.
  144, pp. 111\,608:1--11, Sept. 2020.

\bibitem{Song_Energies_Dec_2019}
H.~Y. Song, G.~S. Lee, and Y.~T. Yoon, ``Optimal operation of critical peak
  pricing for an energy retailer considering balancing costs,'' \emph{MDPI
  Energies}, vol.~12, no.~24, pp. 4658:1--20, Dec. 2019.

\bibitem{Zhang_TSG_Nov_2019}
K.~{Zhang}, S.~{Hanif}, C.~M. {Hackl}, and T.~{Hamacher}, ``A framework for
  multi-regional real-time pricing in distribution grids,'' \emph{IEEE
  Transactions on Smart Grid}, vol.~10, no.~6, pp. 6826--6838, Nov. 2019.

\bibitem{JAn_AE_Mar_2020}
J.~An, M.~Lee, S.~Yeom, and T.~Hong, ``{Determining the Peer-to-Peer
  electricity trading price and strategy for energy prosumers and consumers
  within a microgrid},'' \emph{Applied Energy}, vol. 261, pp. 114\,335:1--16,
  Mar. 2020.

\bibitem{YJiang_AE_Aug_2020}
Y.~Jiang, K.~Zhou, X.~Lu, and S.~Yang, ``{Electricity trading pricing among
  prosumers with game theory-based model in energy blockchain environment},''
  \emph{Applied Energy}, vol. 271, pp. 115\,239:1--16, Aug. 2020.

\bibitem{Anees_AE_Nov_2019}
A.~Anees, T.~Dillon, and Y.-P.~P. Chen, ``{A novel decision strategy for a
  bilateral energy contract},'' \emph{Applied Energy}, vol. 253, pp.
  113\,571:1--113\,571:13, Nov. 2019.

\bibitem{Wang_AE_Nov_2019}
Y.~Wang, K.~Lai, F.~Chen, Z.~Li, and C.~Hu, ``{Shadow price based co-ordination
  methods of microgrids and battery swapping stations},'' \emph{Applied
  Energy}, vol. 253, pp. 113\,510:1--113\,510:16, Nov. 2019.

\bibitem{Yildiz_AE_Dec_2017}
B.~Yildiz, J.~I. Bilbao, J.~Dore, and A.~B. Sproul, ``{Recent advances in the
  analysis of residential electricity consumption and applications of smart
  meter data},'' \emph{Applied Energy}, vol. 208, pp. 402--427, Dec. 2017.

\bibitem{Chen_TSG_July_2019}
T.~{Chen} and W.~{Su}, ``Indirect customer-to-customer energy trading with
  reinforcement learning,'' \emph{IEEE Transactions on Smart Grid}, vol.~10,
  no.~4, pp. 4338--4348, July 2019.

\bibitem{Jogunola_Energies_Dec_2017}
O.~Jogunola, A.~Ikpehai, K.~Anoh, B.~Adebisi, M.~Hammoudeh, S.-Y. Son, and
  G.~Harris, ``State-of-the-art and prospects for peer-to-peer
  transaction-based energy system,'' \emph{MDPI Energies}, vol.~10, no.~12, pp.
  62:1--62:20, Dec. 2017.

\bibitem{Azim_AE_Apr_2020}
M.~I. Azim, W.~Tushar, and T.~K. Saha, ``{Investigating the impact of P2P
  trading on power losses in grid-connected networks with prosumers},''
  \emph{Applied Energy}, vol. 263, pp. 114\,687:1--12, Apr. 2020.

\bibitem{Nikolaidis_TPWRS_Early_2019}
A.~{Nikolaidis}, C.~A. {Charalambous}, and P.~{Mancarella}, ``A graph-based
  loss allocation framework for transactive energy markets in unbalanced radial
  distribution networks,'' \emph{IEEE Transactions on Power Systems}, vol.~34,
  no.~5, pp. 4109--4118, Sept. 2019.

\bibitem{Baroche_TPWRS_July_2019}
T.~{Baroche}, P.~{Pinson}, R.~L.~G. {Latimier}, and H.~B. {Ahmed}, ``Exogenous
  cost allocation in peer-to-peer electricity markets,'' \emph{IEEE
  Transactions on Power Systems}, vol.~34, no.~4, pp. 2553--2564, July 2019.

\bibitem{Xu_TIE_Nov_2019}
Y.~{Xu}, H.~{Sun}, and W.~{Gu}, ``A novel discounted min-consensus algorithm
  for optimal electrical power trading in grid-connected {DC} microgrids,''
  \emph{IEEE Transactions on Industrial Electronics}, vol.~66, no.~11, pp.
  8474--8484, Nov. 2019.

\bibitem{Thomas_TPS_Early_2018}
T.~{Morstyn} and M.~{McCulloch}, ``Multi-class energy management for
  peer-to-peer energy trading driven by prosumer preferences,'' \emph{IEEE
  Transactions on Power Systems}, vol.~34, no.~5, pp. 4005--4014, Sept. 2019.

\bibitem{Thomas_TSG_July_2020}
T.~{Morstyn}, A.~{Teytelboym}, C.~{Hepburn}, and M.~D. {McCulloch},
  ``Integrating p2p energy trading with probabilistic distribution locational
  marginal pricing,'' \emph{IEEE Transactions on Smart Grid}, vol.~11, no.~4,
  pp. 3095--3106, July 2020.

\bibitem{Zhang_AE_June_2018}
C.~Zhang, J.~Wu, Y.~Zhou, M.~Cheng, and C.~Long, ``{Peer-to-Peer energy trading
  in a Microgrid},'' \emph{Applied Energy}, vol. 220, pp. 1--12, June 2018.

\bibitem{Dahraie_Systems_EA_2020}
\BIBentryALTinterwordspacing
M.~{Vahedipour-Dahraie}, H.~{Rashidizadeh-Kermani}, M.~{Shafie-Khah}, and
  P.~{Siano}, ``Peer-to-peer energy trading between wind power producer and
  demand response aggregators for scheduling joint energy and reserve,''
  \emph{IEEE Systems Journal}, 2020, early access. [Online]. Available:
  \url{https://doi.org/10.1109/JSYST.2020.2983101}
\BIBentrySTDinterwordspacing

\bibitem{Baros_TPWRS_Nov_2017}
S.~{Baros} and M.~D. {Ili\'{c}}, ``Distributed torque control of deloaded wind
  dfigs for wind farm power output regulation,'' \emph{IEEE Transactions on
  Power Systems}, vol.~32, no.~6, pp. 4590--4599, Nov. 2017.

\bibitem{Arsoon_AE_Mar_2020}
M.~M. Arsoon and S.~M. Moghaddas-Tafreshi, ``{Peer-to-peer energy bartering for
  the resilience response enhancement of networked microgrids},'' \emph{Applied
  Energy}, vol. 261, pp. 114\,687:1--14, Mar. 2020.

\bibitem{Ruotsalainen_ERSS_Dec_2017}
J.~Ruotsalainen, J.~Karjalainen, M.~Child, and S.~Heinonen, ``{Culture, values,
  lifestyles, and power in energy futures: A critical peer-to-peer vision for
  renewable energy},'' \emph{Energy Research and Social Science}, vol.~34, pp.
  231--239, Dec. 2017.

\bibitem{LeiLi_RSER_Apr_2019}
L.~Li, H.~Manier, and M.-A. Manier, ``{Hydrogen supply chain network design: An
  optimization-oriented review},'' \emph{Renewable and Sustainable Energy
  Review}, vol. 103, pp. 342--360, Apr. 2019.

\bibitem{Robledo_AE_Apr_18}
C.~B. Robledo, V.~Oldenbroek, F.~Abbruzzese, and A.~J.~M. van Wijk,
  ``{Integrating a hydrogen fuel cell electric vehicle with vehicle-to-grid
  technology, photovoltaic power and a residential building},'' \emph{Applied
  Energy}, vol. 215, pp. 615--629, Apr. 2018.

\bibitem{Zhu_AE_Aug_2020}
D.~Zhu, B.~Yang, Q.~Liu, K.~Ma, S.~Zhu, C.~Ma, and X.~Guan, ``{Energy trading
  in microgrids for synergies among electricity, hydrogen and heat networks},''
  \emph{Applied Energy}, vol. 272, pp. 115\,225:1--14, Aug. 2020.

\bibitem{Hasan_RE_Aug_2020}
H.~Mehrjerdi, ``{Peer-to-peer home energy management incorporating hydrogen
  storage system and solar generating units},'' \emph{Renewable Energy}, vol.
  156, pp. 183--192, Aug. 2020.

\bibitem{Xiao_TPWRS_July_2018}
Y.~{Xiao}, X.~{Wang}, P.~{Pinson}, and X.~{Wang}, ``A local energy market for
  electricity and hydrogen,'' \emph{IEEE Transactions on Power Systems},
  vol.~33, no.~4, pp. 3898--3908, July 2018.

\bibitem{Zhang_TIA_July_2016}
J.~{Zhang}, K.~{Li}, M.~{Wang}, W.~{Lee}, H.~{Gao}, C.~{Zhang}, and K.~{Li},
  ``A bi-level program for the planning of an islanded microgrid including
  caes,'' \emph{IEEE Transactions on Industry Applications}, vol.~52, no.~4,
  pp. 2768--2777, July 2016.

\bibitem{TeMiX2020}
``{TeMiX},'' \url{http://temix.com/}, accessed: 2020-07-13.

\bibitem{Yeloha2020}
``{Yeloha},''
  \url{https://www.crunchbase.com/organization/yeloha#section-overview},
  accessed: 2020-07-13.

\bibitem{PowerNet2020}
{Power Ledger}, ``{American PowerNet, United States - Trading of rooftop solar
  energy},'' \url{https://www.powerledger.io/project/american-powernet/},
  accessed: 2020-07-13.

\bibitem{ShareAndCharge2020}
{SHARE\&CHARGE}, ``{Open charing network - The next level of OCPI-based
  e-roaming},'' \url{https://shareandcharge.com/}, accessed: 2020-07-13.

\bibitem{PeerEnergyCloud2020}
``{Peer Energy Cloud},''
  \url{http://software-cluster.org/projects/peer-energy-cloud/}, accessed:
  2020-07-13.

\bibitem{sonnen2020}
{Sonnen}, ``{It is time to declare your independence},''
  \url{https://sonnengroup.com/sonnencommunity/}, accessed: 2020-07-13.

\bibitem{Powerpeer2020}
``{Powerpeer - Power to the people},'' \url{https://www.powerpeers.nl/login},
  accessed: 2020-07-13.

\bibitem{vendebron2020}
``{vandebron},'' \url{https://vandebron.nl/}, accessed: 2020-07-13.

\bibitem{Empower2017}
E.~{Bullich-Massagu\'{e}}, M.~{Arag\"{u}\'{e}s-Pe\~{n}alba},
  P.~{Olivella-Rosell}, P.~{Lloret-Gallego}, J.~{Vidal-Clos}, and A.~{Sumper},
  ``Architecture definition and operation testing of local electricity markets.
  the empower project,'' in \emph{International Conference on Modern Power
  Systems (MPS)}, Cluj-Napoca, Romania, June 2017, pp. 1--5.

\bibitem{Piclo2020}
{Piclo}, ``{Building a smarter energy future},''
  \url{https://piclo.energy/about#whitepaper}, accessed: 2020-07-13.

\bibitem{SmartTest2020}
``{P2P - SmartTest},'' \url{https://www.p2psmartest-h2020.eu/}, accessed:
  2020-07-13.

\bibitem{ARENA_P2P_2020}
{Australian Renewable Energy Agency}, ``{AGL} virtual trial of peer-to-peer
  energy trading,''
  \url{https://arena.gov.au/projects/agl-virtual-trial-peer-to-peer-trading/},
  accessed: 2020-07-13.

\bibitem{RENeW2020}
{Power Ledger}, ``{RENeW Nexus, Australian Government, Australia},''
  \url{https://www.powerledger.io/project/renew-nexus/}, accessed: 2020-07-13.

\bibitem{Nicheliving2020}
------, ``{Niceliving, Australia},''
  \url{https://www.powerledger.io/project/nicheliving/}, accessed: 2020-07-13.

\bibitem{Retail2020}
------, ``{Vicinity, Australia},''
  \url{https://www.powerledger.io/project/vicinity/}, accessed: 2020-07-13.

\bibitem{Powerclub2020}
------, ``{Powerclub, Australia},''
  \url{https://www.powerledger.io/project/powerclub-sonnen/}, accessed:
  2020-07-13.

\bibitem{EastVillage2020}
------, ``{East Village, Australia},''
  \url{https://www.powerledger.io/project/east-village-australia/}, accessed:
  2020-07-13.

\bibitem{GenY2020}
------, ``{Gen Y, Western Australian Government, Australia},''
  \url{https://www.powerledger.io/project/gen-y/}, accessed: 2020-07-13.

\bibitem{Wongan2020}
------, ``{Wongan-Ballidu, Australia},''
  \url{https://www.powerledger.io/project/wongan-ballidu/}, accessed:
  2020-07-13.

\bibitem{EPCSolar2020}
------, ``{EPC Solar Canberra, Australia},''
  \url{https://www.powerledger.io/project/epc/}, accessed: 2020-07-13.

\bibitem{DeHavilland2020}
------, ``{DeHavilland Apartments \& Element47, Australia},''
  \url{https://www.powerledger.io/project/dehavilland/}, accessed: 2020-07-13.

\bibitem{Japan_1_Pilot}
R.~Asseh, ``{Kansai Electric leads study on blockchain use in distributed
  electric supply},''
  \url{https://coingeek.com/kansai-electric-leads-study-blockchain-use-distributed-electric-supply/},
  accessed: 2020-07-13.

\bibitem{Japan_2_Pilot}
{Power Ledger}, ``{KEPCO, Japan - Peer-to-Peer solar power and REC trading},''
  \url{https://www.powerledger.io/project/kepco/}, accessed: 2020-07-13.

\bibitem{Japan_3_Pilot}
------, ``{Sharing energy \& eRex, Japan Peer-to-peer solar power trading},''
  \url{https://www.powerledger.io/project/sharing-energy-erex/}, accessed:
  2020-07-13.

\bibitem{India_1_Pilot}
------, ``{Uttar Pradesh Government, India - Peer-to-peer solar power
  trading},'' \url{https://www.powerledger.io/project/up-government/},
  accessed: 2020-07-13.

\bibitem{India_2_Pilot}
{Solarplaza}, ``{BSES Rajdhani, India - Peer-to-peer solar power trading},''
  \url{https://www.powerledger.io/project/bses-rajdhani/}, accessed:
  2020-07-13.

\bibitem{Thailand_1_Pilot}
{Power Ledger}, ``{Power Ledger P2P Platform Goes Across the Meter with BCPG at
  T77 Precinct, Bangkok},''
  \url{https://medium.com/power-ledger/power-ledger-p2p-platform-goes-across-the-meter-with-bcpg-at-t77-precinct-bangkok-62df5aba3d0a},
  accessed: 2020-07-13.

\bibitem{Thailand_2_Pilot}
------, ``{TDED, Thailand},'' \url{https://www.powerledger.io/project/tded/},
  accessed: 2020-07-13.

\bibitem{SKorea_1_Pilot}
C.~Mu-Hyun, ``{South Korea to trial blockchain electricity market for
  consumers},''
  \url{https://www.zdnet.com/article/south-korea-to-trial-blockchain-electricity-market-for-consumers/},
  accessed: 2020-07-13.

\bibitem{SKorea_2_Pilot}
{ELECTRON}, ``Electron awarded second beis funding to advance electricity
  flexibility trading in south korea with local partner gridwiz,''
  \url{https://www.electron.org.uk/press-releases/electron-awarded-second-beis-funding-to-advance-electricity-flexibility-trading-in-south-korea-with-local-partner-gridwiz},
  accessed: 2020-07-13.

\bibitem{Singapore_1_Pilot}
W.~Thrill, ``Electrify asia {(ELEC)} - a decentralized market place for energy
  in asia,''
  \url{https://hackernoon.com/electrify-asia-elec-a-decentralized-market-place-for-energy-in-asia-f60680dc0bbb},
  accessed: 2020-07-13.

\bibitem{Malaysia_2_Pilot}
{Power Ledger}, ``{SEDA, Malaysia - Peer-to-peer solar power trading},''
  \url{https://www.powerledger.io/project/seda/}, accessed: 2020-07-13.

\bibitem{Rahimi_EM_June_2018}
F.~A. {Rahimi} and S.~{Mokhtari}, ``Distribution management system for the grid
  of the future: A transactive system compensating for the rise in distributed
  energy resources,'' \emph{IEEE Electrification Magazine}, vol.~6, no.~2, pp.
  84--94, June 2018.

\bibitem{Okawa_Negawatt_Nov_2017}
Y.~{Okawa} and T.~{Namerikawa}, ``Distributed optimal power management via
  negawatt trading in real-time electricity market,'' \emph{IEEE Transactions
  on Smart Grid}, vol.~8, no.~6, pp. 3009--3019, Nov. 2017.

\bibitem{Fairley_Spectrum_Oct_2017}
P.~{Fairley}, ``Blockchain world - feeding the blockchain beast if bitcoin ever
  does go mainstream, the electricity needed to sustain it will be enormous,''
  \emph{IEEE Spectr.}, vol.~54, no.~10, pp. 36--59, Oct. 2017.

\end{thebibliography}

\end{document}